\documentclass[lettersize,journal]{IEEEtran}

\usepackage{amssymb,amsmath}
\usepackage{cite}
\usepackage{graphicx}

\usepackage{psfrag}
\usepackage{url}
\usepackage[latin1]{inputenc}
\usepackage[absolute,overlay]{textpos}
\usepackage{gensymb}
\usepackage{cases}
\usepackage[font=footnotesize]{caption}
\usepackage{float}
\usepackage[linesnumbered,ruled,lined]{algorithm2e}
\usepackage{pgf}

\usepackage[nocomma]{optidef}

\usepackage{flushend}

\usepackage{dirtytalk}
\usepackage[printonlyused]{acronym}

\acrodef{MIMO}{multiple-input multiple-output}
\acrodef{mMIMO}{massive multiple-input multiple-output}
\acrodef{AWGN}{additive white Gaussian noise}
\acrodef{IA}{interference alignment}
\acrodef{SNR}{signal-to-noise ratio}
\acrodef{SINR}{signal-to-interference-plus-noise ratio}
\acrodef{CSI}{channel state information}
\acrodef{RL}{reinforcement learning}
\acrodef{5G}{fifth-generation}
\acrodef{6G}{sixth-generation}
\acrodef{ITU}{International Telecommunication Union}
\acrodef{MC}{massive communication}
\acrodef{HRLLC}{hyper-reliable low-latency communication}
\acrodef{VR}{virtual reality}
\acrodef{AR}{augmented reality}
\acrodef{AI}{artificial intelligence}
\acrodef{mMTC}{massive machine type communication}
\acrodef{URLLC}{ultra-reliable low-latency communication}
\acrodef{IA}{interference alignment}
\acrodef{TDMA}{time division multiple access}
\acrodef{FDMA}{frequency division multiple access}
\acrodef{CSI}{channel state information}
\acrodef{i.i.d.}{independent and identically distributed}
\acrodef{NLP}{natural language processing}
\acrodef{RNN}{recurrent neural network}
\acrodef{LSTM}{long short-term memory}
\acrodef{1D}{1-dimensional}
\acrodef{3D}{3-dimensional}
\acrodef{CNN}{convolutional neural network}
\acrodef{ReLU}{rectified linear unit}
\acrodef{tanh}{hyperbolic tangent}
\acrodef{IoE}{internet of everything}
\acrodef{FIR}{finite impulse response}
\acrodef{MSE}{mean squared error}
\acrodef{NMSE}{normalized mean squared error}
\acrodef{EBs}{Exabytes}
\acrodef{ML}{machine learning}
\acrodef{DL}{deep learning}
\acrodef{GRU}{gated recurrent unit}
\acrodef{Tx}{transmitter}
\acrodef{Rx}{receiver}
\acrodef{SVD}{singular value decomposition}
\acrodef{bps/Hz}{bits per second per Hertz}
\acrodef{UHD}{ultra high definition}
\acrodef{MDP}{Markov decision process}
\acrodef{DQN}{deep Q-networks}
\acrodef{DPG}{deterministic policy gradient}
\acrodef{DDPG}{deep deterministic policy gradient}
\acrodef{MADDPG}{multi-agent deep deterministic policy gradient}
\acrodef{DNN}{deep neural network}
\acrodef{MARL}{multi-agent reinforcement learning}
\acrodef{DRL}{deep RL}
\acrodef{V2V}{vehicle to vehicle}
\acrodef{HetNet}{heterogeneous network}
\acrodef{MDP}{Markov decision process}
\acrodef{UAV}{unmanned aerial vehicle}
\acrodef{mmWave}{millimeter-wave}
\acrodef{V2X}{vehicle to everything}
\acrodef{RSMA}{rate-splitting multiple access}
\acrodef{CTE}{centralized training and execution}
\acrodef{CTDE}{centralized training and decentralized execution}
\acrodef{DTE}{decentralized training and execution}
\acrodef{RL-Blind}{blind subspace coordination with RL}
\acrodef{RL-MaxSNR}{maximum-SNR subspace coordination with RL}
\acrodef{Blind-ICISC}{blind inter-cell interference subspace coordination}
\acrodef{MaxSNR-ICISC}{maximum-SNR inter-cell interference subspace coordination}
\acrodef{MMSE}{minimum mean square error}
\acrodef{TF-IA}{transformer-aided IA}
\acrodef{LSTM-IA}{LSTM-aided IA}
\acrodef{CNN-IA}{CNN-aided IA}
\acrodef{WMMSE}{weighted MMSE}
\acrodef{FLOPs}{floating point operations}
\acrodef{FC}{fully connected}
\acrodef{DoF}{degrees-of-freedom}
\usepackage{subcaption}
\usepackage{stfloats}
\usepackage{cuted}

\usepackage{amsthm}

\usepackage{booktabs}
\usepackage{color,soul}

\usepackage{textcomp}
\usepackage{bbm}
\usepackage{xcolor}
\def\BibTeX{{\rm B\kern-.05em{\sc i\kern-.025em b}\kern-.08em
    T\kern-.1667em\lower.7ex\hbox{E}\kern-.125emX}}

\usepackage{tabularx}
\usepackage{makecell}
\usepackage{multirow}

\begin{document}

\title{A Novel Reinforcement Learning Based Framework for Scalable MIMO Interference Alignment}
\author{
	\IEEEauthorblockN{Samitha Gunarathne, \textit{Graduate Student Member, IEEE},  Eslam Eldeeb, \textit{Member, IEEE}, Nurul Huda Mahmood, \textit{Member, IEEE} and Italo Atzeni, \textit{Senior Member, IEEE}
}

    \thanks{The authors are with the Centre for Wireless Communications (CWC), University of Oulu, Finland. (e-mail: samitha.gunarathne@oulu.fi; eslam.eldeeb@oulu.fi; nurulhuda.mahmood@oulu.fi; italo.atzeni@oulu.fi).}

    \thanks{This work was supported by the Research Council of Finland projects Profi6 (grant number: 336449), HIGH-6G (grant number: 348396), 6G Flagship (grant number: 369116), and ReWIN-6G (grant number: 357120), and by the Business Finland's 6GBridge project Local 6G (grant number: 8002/31/2022).}

    \thanks{Copyright (c) 2026 IEEE. Personal use of this material is permitted. However, permission to use this material for any other purposes must be obtained from the IEEE by sending a request to pubs-permissions@ieee.org.}

}
\maketitle

\begin{abstract}

Interference alignment (IA) is a widely recognized approach for mitigating inter-cell interference in multi-user multiple-input multiple-output (MIMO) networks. Despite its effectiveness, practical deployment remains constrained by two major challenges, \emph{i.e.}, the need for global channel state information (CSI) at each transmitter and the complexity of deriving closed-form solutions for intricate MIMO systems. This work aims to maximize network throughput by effectively mitigating interference using an IA-inspired learning algorithm that addresses its aforementioned challenges. First, we propose a predictive, transformer-based IA framework that estimates CSI to reduce signaling overhead in small-scale MIMO systems. Next, we formulate the IA problem as a multi-objective optimization problem based on subspace coordination and develop two reinforcement learning-based algorithms to enhance the scalability of IA in large-scale MIMO systems. Simulation results demonstrate that the proposed methods significantly outperform conventional baselines, with up to $30\%$ average user throughput gains over the best performing baseline.

\end{abstract}

\section{Introduction}\label{sec: introduction}

With the explosive growth of users, data, and applications, \ac{6G} wireless systems aim to transform communications through \ac{AI}-driven automation capable of managing increasingly complex data traffic. The draft framework for \ac{6G} system requirements released by the \ac{ITU} outlines six usage scenarios supported by stringent performance targets, such as extremely high data rates, dense deployment scenarios, ultra high reliability and low latency~\cite{IMT_2030, The_Roadmap_to_6G, AI_Empowered_Multiple_Access_for_6G}. Recent research on \ac{6G} outlines different technology enablers to support these stringent requirements~\cite{rajatheva2020whitepaperbroadbandconnectivity, mahmoodMTC_Eurasip2021, ali20206gwhitepapermachine}. Among these, massive \ac{MIMO} technology is a key enabler for achieving the required spectral efficiency and data rates in \ac{6G} wireless networks. The large number of antennas at the base stations in massive \ac{MIMO} provides seamless connectivity to the end users and highly directional beamforming capabilities~\cite{survey_on_ML_for_mMIMO}. Incorporating massive \ac{MIMO} into ultra-dense networks enhances connectivity for highly mobile users and extends coverage over large geographical areas~\cite{6G_Wireless_System, survey_on_ML_for_mMIMO, Ultra-Dense_Net_B5G:_Heterogeneous_Moving_Cells}. Aggressive spatial multiplexing in massive \ac{MIMO} can increase system capacity nearly tenfold and the energy efficiency by up to two orders of magnitude~\cite{mMIMO_for_Next_Gen_Wireless_Systems}. 

However, the ultra-dense deployment of base stations and user devices makes interference a major challenge in \ac{MIMO} systems. Therefore, interference management has become an active research area in the field of wireless communications. Several approaches have been proposed to mitigate inter-cell interference in wireless networks, including \ac{IA}, interference cancellation, interference coordination, beamforming, and coordinated multipoint transmission. This work particularly focuses on \ac{IA}, which is a \ac{DoF}-optimal interference management technique that approaches the capacity of an interference network in the regime of high \ac{SNR}~\cite{IA-3}. \Ac{IA} schemes ensure that all interference signals at unintended receivers are aligned within the same subspace, thereby minimizing the dimension of the interference subspace. The desired signal is then recovered using a carefully designed post-processing matrix that suppresses this interference. 

Despite its theoretical effectiveness, the \ac{IA} concept has several practical limitations. First, it requires global \ac{CSI} at each node, which is used to compute the appropriate precoding matrices. Second, it does not scale well as the number of users in the network increases. To address the global \ac{CSI} requirement, several studies have proposed iterative algorithms that exploit channel reciprocity to achieve \ac{IA} using only local \ac{CSI} at each node, thereby reducing the dependence on global \ac{CSI}~\cite{Distributed_Numerical_Approach_to_IA}. In addition, centralized feedback topologies have been introduced for \ac{MIMO} interference channels under \ac{IA} constraints to reduce \ac{CSI} overhead by supporting sequential \ac{CSI} exchange~\cite{IA-2}.

In this study, we consider both small-scale and large-scale \ac{MIMO} systems to evaluate the scalability of \ac{IA} approaches. As the number of users increases, \ac{IA} becomes difficult to scale due to its rapidly growing complexity. This occurs because each additional user increases the dimensionality of the \ac{IA}, thereby enlarging the precoding matrices and the number of alignment constraints that must be satisfied. For instance, if the users are more than three, it is nontrivial to derive a closed-form solution using \ac{IA}~\cite{low_compexity_IA,Linear_Transceiver_Design_for_Interference_Alignment, Interference_Alignment_as_a_Rank_Constrained_Rank_Minimization}. Therefore, calculating optimal precoders and post-processing matrices becomes a non-convex and NP-hard problem for complex settings~\cite{low_compexity_IA, Linear_Transceiver_Design_for_Interference_Alignment}. 
This motivates the development of scalable interference management schemes that efficiently project both desired and interference signals onto their respective subspaces at the receiver in large-scale \ac{MIMO} systems.

To address the requirement for global \ac{CSI}, we propose a method that predicts the global \ac{CSI} from the previous channel data and proactively applies \ac{IA}. This approach reduces the \ac{CSI} overhead, making \ac{IA} a practically feasible interference management solution for small-scale \ac{MIMO} systems. The \ac{CSI} prediction challenge is formulated as a time-series forecasting problem, where we propose to adopt the transformer architectures due to its superior performance over other time-series models, such as \ac{RNN} and \ac{LSTM}. Transformers effectively resist the well-known vanishing gradient problem commonly associated with \acp{RNN}~\cite{RNN_vanishing_gradient}. Additionally, unlike \acp{LSTM}, transformers can process data in parallel, making them more efficient in training large datasets. Given their capacity to capture global contextual information, transformer models are especially well-suited for \ac{CSI} prediction.

The scalability challenge is addressed by adopting a \ac{DRL} framework for optimizing the \ac{IA} precoding and post-processing matrices. \Ac{RL} offers a suitable framework, enabling agents to sense the environment and make optimal decisions through interaction. As a model-free method, decision making in \ac{RL} occurs without prior knowledge of the environment. \Ac{DRL} extends this capability by integrating \acp{DNN} with \ac{RL}, providing high efficiency in complex, high-dimensional settings. Moreover, \ac{DRL} is well suited for solving NP-hard problems.

Primary findings of this work was reported in~\cite{TF_IA_EUCNC}, which primarily focused on reducing the \ac{CSI} feedback overhead of \ac{IA} through transformer-aided \ac{CSI} prediction. Since the work in~\cite{TF_IA_EUCNC} is limited to \ac{IA} in small-scale \ac{MIMO} systems, the present study generalizes the system model, reformulates the sum throughput maximization problem as a distance minimization problem, and introduces an \ac{RL}-based subspace coordination method to enhance \ac{IA} in large-scale \ac{MIMO} systems.

\subsection{Motivation  and Contributions} \label{subsec: contributions}

We propose two complementary approaches to address the practical limitations of \ac{IA}, \emph{i.e.}, the need for global \ac{CSI} at the transmitter and limited scalability with a large number of users. To the best of our knowledge, no prior work has investigated integrating \ac{CSI} prediction into \ac{IA} to reduce \ac{CSI} overhead. This study fills that gap by introducing a transformer-based framework for accurate \ac{CSI} forecasting, which is then used to compute precoding matrices in a small-scale multi-user \ac{MIMO} system, effectively mitigating the \ac{CSI} overhead problem. Regarding scalability, as the number of transmitter-receiver pairs increases, the complexity of \ac{IA} grows exponentially~\cite{IA-8, IA-9}. The limited scalability of \ac{IA} motivates the adoption of learning-based interference management instead of closed-form solutions. To overcome this, we propose a \ac{RL}-based scheme to enhance \ac{IA} performance in large-scale \ac{MIMO} systems while maintaining efficiency. The main contributions of this study are summarized as follows:
\begin{itemize}
    \item We conduct a comprehensive analysis of the factors influencing \ac{IA} in a multi-user \ac{MIMO} system and identify key parameters that are optimized to enable an efficient interference management scheme.
    
    \item We propose a novel method, termed \textit{\ac{TF-IA}}, which employs a \ac{CSI} prediction technique to reduce signaling overhead in \ac{IA} for small-scale \ac{MIMO} systems and demonstrate its superior performance compared to two baseline methods.
    
    \item We formulate a multi-objective optimization problem based on subspace coordination to enhance scalability of \ac{IA}. We introduce two \ac{RL}-based models, \emph{i.e.}, \textit{\ac{RL-Blind}}, which assigns the subspaces arbitrarily, and \textit{\ac{RL-MaxSNR}}, which assigns them via eigendecomposition of the channel matrix.
    
    \item We evaluate the performance of the proposed methods in small-scale and large-scale \ac{MIMO} systems, where simulation results demonstrate that our approaches outperform conventional ones in terms of sum rate and average user throughput.
\end{itemize}

\subsection{Notations and Structure of the Paper} \label{subsec: notations}

Vectors and matrices are denoted by bold lowercase and uppercase letters, respectively. The $i$-th entry of a vector $\mathbf{x}$ is denoted by $\mathbf{x}_{i}$, while the element in the $i$-th row and $j$-th column of a matrix $\mathbf{X}$ is denoted by $\mathbf{X}_{i,j}$. The operators $(\cdot)^{\mathrm{T}}$ and $(\cdot)^{\mathrm{H}}$ represent transpose and conjugate transpose of a matrix, respectively. The $n$-dimensional identity matrix and all-one vector are denoted by $\mathbf{I}_n$ and $\mathbf{1}_n$, respectively. The sets of complex and real numbers are represented by $\mathbb{C}$ and $\mathbb{R}$, respectively. The real and imaginary parts of a complex number are denoted by $\textrm{Re}[\cdot]$ and $\textrm{Im}[\cdot]$, respectively. The notation $\mathcal{CN}(0,\sigma^{2})$ denotes a complex Gaussian distribution with zero mean and variance $\sigma^{2}$. The Frobenius norm is denoted by $\|\cdot\|_{\textrm{F}}$. The expectation operator is represented by $\mathbb{E}\{\cdot\}$, whereas $\nabla(\cdot)$ denotes the gradient.

 The remainder of this paper is organized as follows. Section~\ref{sec: prior_work} summarizes the related prior works. Section~\ref{sec: multi_user_system_model} describes the considered multi-user \ac{MIMO} system model, along with the fundamentals of \ac{IA} and the problem formulation. Section~\ref{sec: TF_IA} presents the proposed transformer-based \ac{CSI} prediction method for \ac{IA}. The \ac{RL}-based subspace coordination approach for \ac{IA} is detailed in Section~\ref{sec: RL_IA}. Simulation results and an in-depth discussion are provided in Section~\ref{sec: simulation_results}. Lastly, Section~\ref{sec: Conclusion} concludes the paper.

\section{Prior Works} \label{sec: prior_work}

This section reviews prior work on \ac{IA}, \ac{CSI} prediction, and \ac{RL}-based wireless resource management.

\subsection{Studies on IA} \label{subsec: IA literature}

The \Ac{IA} algorithm for \ac{MIMO} communication is initially investigated in~\cite{IA-1}, where two signaling schemes are introduced for a multiple-antenna system with two transmitters and two receivers, considering both transmitter and receiver perspectives. In these schemes, non-linear filters are employed at both ends, enabling the decomposition of the system into non-interfering broadcast or multiple-access channels. The main goal of such implementation is to achieve the highest multiplexing gain. The work in~\cite{Feasibility_of_IA} investigates the feasibility of \ac{IA} in signal vector spaces for $K$-user \ac{MIMO} interference channels by analyzing the resolvability of multivariate polynomial systems. In~\cite{IA_for_MIMO_downlink_multicell_networks}, \ac{IA} is studied for \ac{MIMO} multi-cell downlink networks with square channel matrices, where a joint user grouping and base station association is combined with a design of the transmit and receive beamforming matrices based on generalized eigendecomposition.

An \ac{IA} scheme for multi-cell, multi-user \ac{MIMO} systems under Gaussian interference broadcast channels is proposed in~\cite{IA_Techniques_for_MIMO}. The transmit and receive beamforming vectors are jointly designed using the grouping methods described in prior studies and a multiple-access channel-broadcast channel duality is introduced to perform \ac{IA} while maximizing the performance within each cell. The resulting dual problem is convex and solved using the interior-point method. The work in~\cite{IA_with_Limited_Feedback} presents a quantitative performance analysis of \ac{IA} under limited feedback and derived a closed-form expression for the average transmission rate as a function of \ac{CSI} accuracy and the number of data streams. Furthermore, an adaptive feedback allocation scheme is proposed, and the corresponding theoretical claims are validated through simulation results.

\subsection{Studies on CSI Prediction} \label{subsec: CSI prediction literature}

Acquiring accurate \ac{CSI} is crucial for efficient resource management in wireless communication systems. Consequently, \ac{CSI} prediction has become an active research topic in recent years. The work in~\cite{CSI_pred-1} investigates \ac{ML}-based \ac{CSI} prediction for massive \ac{MIMO}, proposing a three-dimensional complex \ac{CNN} that has been tested with both simulated and field data. The work in~\cite{CSI_pred-3} proposes an \ac{ML}-based time-division duplex scheme that obtains \ac{CSI} by learning temporal correlations of the channel. Specifically, the \ac{CSI} data is represented as images and processed using a \ac{CNN} to extract spatial patterns, followed by an autoregressive predictor for channel forecasting. However, since autoregressive models enable only one-step prediction, the approach is limited in its effectiveness for \ac{CSI} prediction. The work in~\cite{CSI_pred-2} presents \ac{CSI} prediction with \ac{RNN} incorporating \ac{LSTM} and \ac{GRU}, mainly focusing on rapid channel variation due to multi-path channel fading. The work in~\cite{CSI_pred-4} explores the potential of transformer-based channel prediction, testing the proposed framework with real-world data. Therein, \ac{CSI} prediction is performed using both the encoder and decoder modules of the transformer architecture, which increases model complexity for time-series forecasting tasks. 

In~\cite{CSI_Prediction_for_5G_using_DL}, an online \ac{CSI} prediction scheme that leverages historical data is proposed. The framework is designed by incorporating \ac{CNN} and \ac{LSTM} architectures, and is complemented by an offline-online two-step training mechanism to achieve stable results. The work in~\cite{Transformer-Empowered_Predictive_Beamforming} proposes a technique based on deep learning to predict the precoder design from historical \ac{CSI} data. To implement this design, transformers and \ac{CNN} architectures are utilized to extract spatial-temporal features from previous \ac{CSI} data. Both~\cite{Transformer-Based_Channel_Prediction_for_Rate-Splitting_Multiple_Access, Transformer-Empowered_Predictive_Beamforming} employ encoder-decoder architectures for \ac{CSI} prediction, although the decoder component is not essential for time-series forecasting. Most existing \ac{CSI} prediction studies have focused on various interference management approaches; however, the application of \ac{CSI} prediction for \ac{IA} has not been adequately explored.

\subsection{Studies on RL-Based Wireless Resource Management} \label{subsec: RL-based literature}

In recent years, \ac{RL}-based approaches have shown strong potential in addressing the challenges of resource management in wireless communication networks. For instance,~\cite{MARL_in_wireless_resource_management} proposes a distributed resource management framework with interference mitigation for wireless networks based on \ac{MARL}. Each transmitter acts as a deep \ac{RL} agent and jointly optimizes power control and user selection in a decentralized manner. The work in~\cite{eldeeb2022multi} investigates the role of \ac{RL} in \ac{UAV} trajectory planning to minimize power consumption while maximizing information freshness. In~\cite{Spectrum_sharing_MARL}, spectrum sharing in vehicular networks incorporating deep \ac{MARL} is investigated, where each link is modeled as an autonomous agent that interacts with the communication environment and learns to optimize spectrum sharing and power allocation through Q-networks. The work in~\cite{Joint_IA_and_power_allocation_using_DRL} proposes an interference suppression scheme for heterogeneous networks by applying co-tier intra-cell \ac{IA} and formulating power control problem as a \ac{MDP}, which is solved using a \ac{DDPG}-based algorithm. However, since the approach still relies on \ac{IA} in conjunction with \ac{RL}, the requirement for global \ac{CSI} remains a limitation.
 
The work in~\cite{RL_with_selective_exploration_for_IM} addresses the joint optimization of beamforming, power control, and interference management in multi-cell millimeter-wave networks, proposing both single- and multi-agent \ac{RL} approaches for centralized and distributed settings, respectively. Although this framework eliminates the need for explicit \ac{CSI}, the multi-agent configuration introduces higher computational complexity. In~\cite{RL-Based_Downlink_Interference_Control_for_Ultra-Dense_Small_Cells}, a \ac{RL}-based power control framework is proposed to suppress inter-cell interference and improve energy efficiency in ultra-dense small-cell downlink deployments. Recent studies have demonstrated that \ac{RL}-based interference management achieves notable performance gains in dynamic wireless environments. However, these results depend on having global \ac{CSI}, which limits their use in real-world situations.

\section{System Model} \label{sec: multi_user_system_model}

This section presents the details of the multi-user \ac{MIMO} communication model, the fundamentals of \ac{IA}, and the problem formulation.

\subsection{Multi-User MIMO System Model} \label{subsec: multi user system model}

\begin{figure}[t!]
    \centering
    \includegraphics[width=1\columnwidth,trim={0 0 0 0},clip]{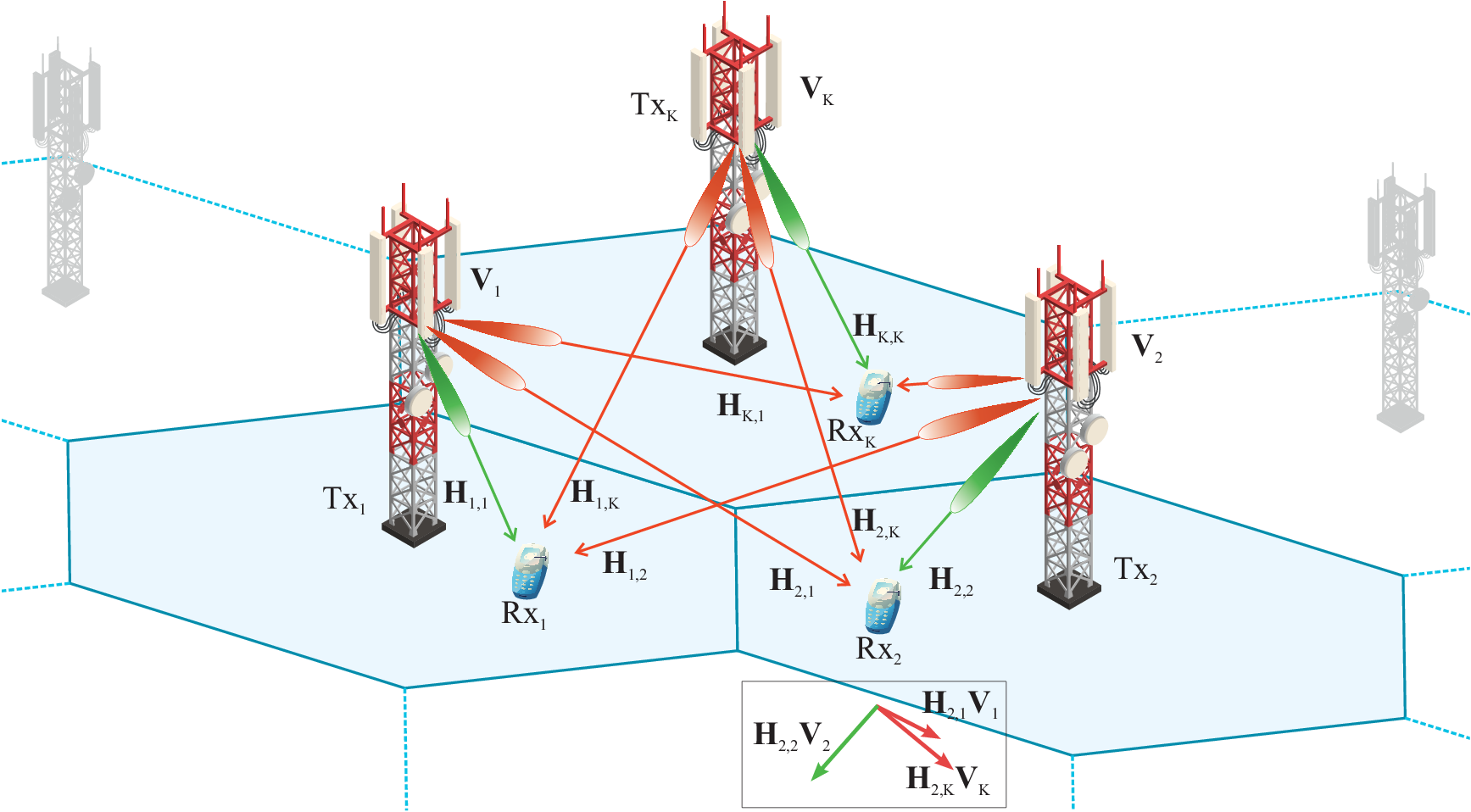}
    \caption{Multi-user \ac{MIMO} system model with $K$ transmitters and receivers.}
    \label{fig: System_diagram}
\end{figure}

Consider a downlink $K$-user \ac{MIMO} interference network, where each transmitter and receiver are equipped with $m_{\textrm{t}}$ transmit antennas and $n_{\textrm{r}}$ receive antennas, respectively. A set of $d_{j} \leq \min(m_{\textrm{t}}, n_{\textrm{r}})$ data streams per user is transmitted at each transmission block. This wireless channel is statistically modeled as a time-correlated static Rayleigh fading \ac{MIMO} channel. All desired transmissions occur between pairs with the same index, \emph{i.e.}, the $k$-th transmitter communicates with the $k$-th receiver. Consequently, all other received signals are considered as interference. The received signal at the $j$-th user and $t$-th transmission interval ($\mathbf{y}_{j}(t) \in \mathbb{C}^{n_{\textrm{r}} \times 1}$) can be expressed as
\begin{align} 
\label{eq: received signal}
    \mathbf{y}_{j}(t) = \ & \sqrt{\alpha_{j,j}\frac{P}{d_{j}}} \: \mathbf{H}_{j, j}(t)\mathbf{V}_{j}(t) \: \mathbf{x}_{j}(t) \nonumber \\ & + \sum_{i \neq j} \sqrt{\alpha_{j, i}\frac{P}{d_{i}}} \: \mathbf{H}_{j, i}(t)\mathbf{V}_{i}(t) \: \mathbf{x}_{i}(t) + \mathbf{z}_{j}(t),  
\end{align}
where $P$ and $\alpha_{i,j}$ denote, respectively, the transmit power and the path loss factor corresponding to the long-term fading between the $i$-th receiver and $j$-th transmitter. $\mathbf{H}_{j, i} \in \mathbb{C}^{n_{\textrm{r}} \times m_{\textrm{t}}}$ and $\mathbf{V}_{j} \in \mathbb{C}^{m_{\textrm{t}} \times d_{j}}$ denote the channel coefficients matrix between the $i$-th transmitter and $j$-th receiver and the precoding matrix of the $j$-th transmitter with orthonormal columns ($\mathbf{V}^{\mathrm{H}}_{j} \mathbf{V}_{j} = \mathbf{I}_{d_{j}}, \forall j$), respectively. $\mathbf{x}_j(t) \in \mathbb{C}^{d_j \times 1}$ represents the vector containing the $d_{j}$ independently encoded data streams transmitted from the $j$-th transmitter and $\mathbf{z}_{j}(t)$ denotes the \ac{AWGN} vector at the $j$-th receiver with \ac{i.i.d.} $\mathcal{CN}(0,\sigma_{\textrm{w}}^{2})$ elements. We also assume that the elements of $\mathbf{H}_{j, i}$ and $\mathbf{x}_{i}$ are \ac{i.i.d.}. The time index $t$ is dropped henceforth for ease of presentation. The aforementioned system model will be used throughout this study and is depicted in Fig.~\ref{fig: System_diagram}.

\subsection{Fundamentals of IA} \label{subsec: IA fundamentals}

The above \ac{MIMO} system can be viewed differently depending on the perspective. From the transmitters' point of view, the system functions as $K$ broadcast channels~\cite{IA-1}. Conversely, from the receivers' perspective, the system operates as $K$ multiple access channels~\cite{IA-1}. The precoding matrices $\mathbf{V}_{i}$ at all transmitters are designed to ensure that all interference signal subspaces are aligned into a single subspace, while the desired signal subspace remains linearly independent of the interfering subspaces. As a result, each receiver can employ the zero-forcing technique to mitigate interference signals while preserving the desired signal. The following two conditions must be satisfied to achieve \ac{IA}:
\begin{align} \label{eq: IA eq 1}
\mathbf{U}_{j}^{\mathrm{H}}\mathbf{H}_{j,i}\mathbf{V}_{i} & = \mathbf{0}, \quad \forall i \neq j, \\
\label{eq: IA eq 2}
\mathrm{rank}(\mathbf{U}_{j}^{\mathrm{H}}\mathbf{H}_{j,j}\mathbf{V}_{j}) & = d_{j}, \quad \forall j \in \{1, \dots, K\},
\end{align}
where $\mathbf{U}_{j}$ is the post-processing zero-forcing filters at the $j$-th receiver~\cite{IA-4}. The \ac{IA} technique involves designing both the transmit and receive matrices at the transmitter, which requires global \ac{CSI} and a feedback mechanism to communicate the receive matrices to the receivers. When these two requirements are met, the \ac{IA} conditions in \eqref{eq: IA eq 1} and \eqref{eq: IA eq 2} ensure that each transmitter $j$ can design linear transmit and receive strategies to deliver $d_{j}$ independent data streams to the $j$-th receiver without interference~\cite{IA-2}.

Due to antenna correlation and imperfect alignment, a whitening process is required to restore the \ac{AWGN} model. To design a filter that whitens the interference plus noise component, the total interference covariance matrix for the $j$-th receiver is computed as
\begin{equation} \label{eq: interference_covariance}
    \mathbf{\Phi}_{j} =  \sum_{i \neq j}\mathbf{H}_{j,i}\mathbf{V}_{i}\mathbf{V}^{\mathrm{H}}_{i}\mathbf{H}^{\mathrm{H}}_{j,i} + \mathbf{I}_{n_{\textrm{r}}},
\end{equation}
where $\mathbf{I}_{n_{\textrm{r}}}$ is introduced to ensure numerical stability. The whitening filter is defined as $\mathbf{\Psi}_{j} = \mathbf{\Phi}_{j}^{-1/2}$ and applied to $\mathbf{H}_{j,j}$, yielding the transformed channel matrix $\hat{\mathbf{H}}_{j,j} = \mathbf{\Psi}_{j}\mathbf{H}_{j,j}$. The precoding and post-processing matrices are derived using the scheme proposed in~\cite{An_efficient_signaling_scheme_for_MIMO_Maddah_Ali}. The columns of the precoding matrix $\mathbf{V}_{j}$ correspond to the $d_{j}$ right singular vectors associated with the $d_{j}$ largest singular values of $\hat{\mathbf{H}}_{j,j}$. Then, the $i$-th column of $\mathbf{U}_{j}$ can be expressed as~\cite{An_efficient_signaling_scheme_for_MIMO_Maddah_Ali}
\begin{equation} \label{eq: ith_column_of_Uj}
    \mathbf{u}_{j}^{(i)} =  \frac{\hat{\mathbf{H}}_{j,j}\mathbf{v}_{j}^{(i)}}{\lambda_{j}^{(i)}},
\end{equation}
where $\mathbf{v}_{j}^{(i)}$ denotes the $i$-th column of $\mathbf{V}_{j}$ and $\lambda_{j}^{(i)} = \| \hat{\mathbf{H}}_{j,j}\mathbf{v}_{j}^{(i)} \|$. To perform the alignment, perfect global \ac{CSI} at the transmitter is required to compute the precoding and post-processing matrices. This incurs significant signaling overhead, which is one of the main challenges of implementing \ac{IA} in practical systems. In this work, we address this issue by employing efficient \ac{CSI} acquisition methods at the transmitter, as detailed in Section~\ref{sec: TF_IA}.

\subsection{Problem Formulation} \label{subsec: problem statement}

According to the system model described in Section~\ref{subsec: multi user system model}, the received signal $\mathbf{y}_{j}$ contains both interference and noise. The primary motivation for applying \ac{IA} in \ac{MIMO} systems is to maximize the network throughput by mitigating interference. In particular, the maximum sum throughput can be formulated as an optimization problem in which the precoders are designed according to the principles of \ac{IA}. Following~\cite{Inter_Cell_Interference_Sub_Space_Coordination}, this optimization problem is expressed as
\begin{maxi!}|l|[2] 
{\mathbf{V}_{j}}{\sum_{j=1}^{K} R_{j}}
{}{\mathbb{P}1: \quad}
{\label{eq: tp_maximization_objective}}
\addConstraint{\mathbf{V}^{\mathrm{H}}_{j} \mathbf{V}_{j} = \mathbf{I}_{d_{j}}, \quad \forall j \in \{1, \dots, K\}},
\end{maxi!}
where $R_{j}$ denotes the rate at the $j$-th receiver after applying the \ac{MMSE} filter given by
\begin{equation} \label{eq: throughput}
R_{j} = \mathrm{logdet} \left( \mathbf{I}_{n_{\textrm{t}}} + \mathbf{G}_{j,j}^{\mathrm{H}} \left( \sum_{i \neq j}^{K} \mathbf{G}_{j,i}\mathbf{G}_{j,i}^{\mathrm{H}} + \mathbf{I}_{n_{\textrm{r}}} \right) ^{-1} \mathbf{G}_{j,j} \right).
\end{equation}
Here $\mathbf{G}_{j,j} = \sqrt{\alpha_{j, j}\frac{P}{d_{j}}} , \mathbf{H}_{j,j}\mathbf{V}_{j}$ denotes the effective channel between the $j$-th receiver and its corresponding transmitter~\cite{Inter_Cell_Interference_Sub_Space_Coordination, MMSE_Beamforming_Design_for_a_MIMO_Interference_Channel}.

Problem $\mathbb{P}1$ is non-convex~\cite{uplink_MIMO_transmission_scheme_in_a_multicell_environment} and a closed-form solution is not attainable. Several methods have been proposed in the literature to solve this problem, including alternating optimization~\cite{IA_via_alternating_minimization}, manifold optimization~\cite{IA_manifold_optimization}, and weighted \ac{MMSE}~\cite{IA_WMMSE}. However, alternating optimization is sensitive to initialization and may converge to a poor local optimum, whereas manifold optimization is complex to implement~\cite{practical_chalenges_of_IA}. Moreover, all three methods require global \ac{CSI}. These limitations motivate the development of the two complementary approaches presented in this paper: a predictive \ac{CSI}-based method for proactively applying \ac{IA} in small-scale \ac{MIMO} networks and an \ac{RL}-driven interference management scheme for large-scale \ac{MIMO} systems.

\section{Transformer-Aided IA for Small-Scale MIMO} \label{sec: TF_IA}

This section provides a detailed description of the proposed \ac{TF-IA} scheme, which offers a systematic solution to address the full gloabl \ac{CSI} requirement at the transmitter. First, we introduce the preliminaries of transformers, which are required to present the proposed \ac{TF-IA} algorithm.

\subsection{Preliminaries on Transformers} \label{subsec: Preliminaries_of_TF}

The original vanilla transformer~\cite{Attention_is_all_you_need} is a sequence-to-sequence model composed of an encoder and a decoder, each implemented as a stack of identical blocks. Formally, the encoder implements a mapping $f_{\textrm{enc}} : (x_{1}, \dots, x_{n}) \mapsto (z_{1}, \dots, z_{n})$, where $(z_{1}, \dots, z_{n})$ are continuous latent representations of the input sequence. The decoder then applies a mapping $f_{\textrm{dec}} : (z_{1}, \dots, z_{n}) \mapsto (y_{1}, \dots, y_{m})$, generating the output sequence in an autoregressive manner. The complete transformer architecture is constructed by stacking attention mechanisms and point-wise fully connected layers within each encoder and decoder block.

The attention mechanism in the transformer is defined as a function of the query $\mathbf{Q} \in \mathbb{R}^{N \times D_{\textrm{k}}}$, key $\mathbf{K} \in \mathbb{R}^{M \times D_{\textrm{k}}}$, and value $\mathbf{V} \in \mathbb{R}^{M \times D_{\textrm{v}}}$, where $N $ and $M$ denote the lengths of the query and key sequences, respectively, and $D_{\textrm{k}}$ and $D_{\textrm{v}}$ denote the dimensions of the keys and values, respectively. Using a matrix formulation, the scaled dot-product attention is expressed as~\cite{A_Survey_of_Transformers}
\begin{equation} \label{eq: attention}
\mathrm{Attention}(\mathbf{Q}, \mathbf{K}, \mathbf{V}) = \mathrm{softmax}\left( \frac{\mathbf{Q} \mathbf{K}^{\mathrm{T}}}{\sqrt{D_{\textrm{k}}}} \right),
\end{equation} 
where $\mathrm{softmax}(\cdot)$ denotes the softmax function. However, transformers employ multi-head attention instead of a single attention mechanism. In multi-head attention, the original dimensions $D_{\textrm{m}}$ of $\mathbf{Q}$, $\mathbf{K}$, and $\mathbf{V}$ are linearly projected $h$ times into  $D_{\textrm{k}}$, $D_{\textrm{k}}$, and $D_{\textrm{v}}$ respectively. The attention function is then applied to each head, producing outputs of dimension $D_{\textrm{v}}$. These outputs are concatenated to form the final result, which can be expressed as~\cite{Attention_is_all_you_need, A_Survey_of_Transformers}
\begin{equation} \label{eq: multihead_attention}
\mathrm{MultiHeadAtten}(\mathbf{Q}, \mathbf{K}, \mathbf{V}) = \mathrm{concat}\left( \textrm{head}_{1}, \dots , \textrm{head}_{h} \right) \mathbf{W}^{\mathbf{O}},
\end{equation} 
where $\textrm{concat}(\cdot)$ denotes the concatenation operation and 
\begin{equation} \label{eq: head}
\mathrm{head}_{i} = \mathrm{Attention}\left( \mathbf{Q}\mathbf{W}_{i}^{\mathbf{Q}}, \mathbf{K}\mathbf{W}_{i}^{\mathbf{K}}, \mathbf{V}\mathbf{W}_{i}^{\mathbf{V}} \right).
\end{equation} 
Here, the projection matrices are defined as $\mathbf{W}_{i}^{\mathbf{Q}} \in \mathbb{R}^{D_{\textrm{m}} \times D_{\textrm{k}}}$, $\mathbf{W}_{i}^{\mathbf{K}} \in \mathbb{R}^{D_{\textrm{m}} \times D_{\textrm{k}}}$, $\mathbf{W}_{i}^{\mathbf{V}} \in \mathbb{R}^{D_{\textrm{m}} \times D_{\textrm{v}}}$, and $\mathbf{W}^{\mathbf{O}} \in \mathbb{R}^{hD_{\textrm{v}} \times D_{\textrm{m}}}$.

Before being processed by the encoder block, the input sequence is projected into an embedding space as
\begin{equation} \label{eq: input_embedding}
\mathbf{E} = \mathbf{X}_{t-L+1:t}\mathbf{W}_{\textrm{e}} + \mathbf{1}_{L} \mathbf{b}_{\textrm{e}}^{\mathrm{T}},
\end{equation}
where $\mathbf{X}_{t-L+1:t} \in \mathbb{R}^{L \times f_{\textrm{in}}}$ is the input data sequence, $\mathbf{W}_{\textrm{e}} \in \mathbb{R}^{f_{\textrm{in}} \times D_{\textrm{m}}}$ is the embedding weight matrix, and $\mathbf{b}_{\textrm{e}} \in \mathbb{R}^{D_{\textrm{m}} \times 1}$ is the embedding bias vector. Here, $L$ and $f_{\textrm{in}}$ denote the training window length and the number of input features, respectively.
The model is trained by minimizing the \ac{MSE} loss, defined as
\begin{equation} \label{eq: mse_loss}
\mathcal{L} = \frac{1}{BHf_{\textrm{in}}}\sum_{b=1}^{B}\|\hat{\mathbf{Y}}^{(b)} - \mathbf{Y}^{(b)}\|_{\mathrm{F}}^{2},
\end{equation}
where $\hat{\mathbf{Y}}^{(b)}$ and $\mathbf{Y}^{(b)}$ represent the predicted and true sequences of the $b$-th sample in a batch of size $B$, respectively, and $H$ denotes the forecast horizon.

\subsection{Proposed TF-IA} \label{subsec: TF_IA}

As described in Section~\ref{sec: introduction}, \ac{IA} requires global \ac{CSI} at each transmitter to compute appropriate precoders and post-processing matrices. The proposed \ac{TF-IA} leverages the transformer's capability for time series forecasting to enable accurate \ac{CSI} prediction. The prediction occurs at the transmitter end and historical \ac{CSI} data is used to learn the channel behavior. We assume that the transmitters learn the historical \ac{CSI} data through transmission of dedicated pilot signals along with the uplink feedback message, thereby avoiding the need of a dedicated instantaneous \ac{CSI} feedback loop. Specifically, in this study, the instantaneous channel matrix represents the \ac{CSI} of interest. We approach the \ac{CSI} prediction problem by treating the channel matrix as a compact set of time-series signals. Consequently, we model this problem as a multivariate time series forecasting task and utilize only the encoder block for the entire \ac{CSI} prediction.

At the $t$-th time sample, let the $k$-th column of the channel matrix $\mathbf{H}_{j,i}(t)$ be denoted by $\mathbf{h}_{j,i}^{(k)}(t)$. After collecting a training channel dataset comprising $N$ time samples, the \ac{CSI} data, \emph{i.e.}, $\mathbf{H}_{j,i}$, are reshaped to serve as input to the transformer model as
\begin{equation} \label{eq: reshaped_channel}
\begin{bmatrix}
\textrm{Re}[\mathbf{h}_{j,i}^{(1)}(t)] & \textrm{Re}[\mathbf{h}_{j,i}^{(1)}(t-1)] & \cdots & \textrm{Re}[\mathbf{h}_{j,i}^{(1)}(t-T)]\\
\vdots & \vdots & \vdots & \vdots \\
\textrm{Re}[\mathbf{h}_{j,i}^{(m_{\textrm{t}})}(t)] & \textrm{Re}[\mathbf{h}_{j,i}^{(m_{\textrm{t}})}(t-1)] & \cdots & \textrm{Re}[\mathbf{h}_{j,i}^{(m_{\textrm{t}})}(t-T)]\\
\textrm{Im}[\mathbf{h}_{j,i}^{(1)}(t)] & \textrm{Im}[\mathbf{h}_{j,i}^{(1)}(t-1)] & \cdots & \textrm{Im}[\mathbf{h}_{j,i}^{(1)}(t-T)]\\
\vdots & \vdots & \vdots & \vdots \\
\textrm{Im}[\mathbf{h}_{j,i}^{(m_{\textrm{t}})}(t)] & \textrm{Im}[\mathbf{h}_{j,i}^{(m_{\textrm{t}})}(t-1)] & \cdots & \textrm{Im}[\mathbf{h}_{j,i}^{(m_{\textrm{t}})}(t-T)]\\
\end{bmatrix}.
\end{equation}
The transformer model is then trained to predict the real and imaginary components of $\mathbf{H}_{j,i}$ until the desired prediction accuracy, measured in terms of \ac{MSE}, is achieved. Subsequently, the predicted \ac{CSI} components are reshaped into channel matrices, and \ac{IA} is applied using~\eqref{eq: IA eq 1}--\eqref{eq: ith_column_of_Uj} to compute the estimated precoding and post-processing matrices $\hat{\mathbf{V}}_{j}$ and $\hat{\mathbf{U}}_{j}$. Finally, the transmission is carried out over the true channel. The \ac{SINR} over all the streams of the $j$-th receiver, obtained using the transmit and receive matrices computed based on the estimated channels, is expressed as~\cite{TF_IA_EUCNC}
\begin{equation} \label{eq: predicted_SINR}
\hat{\gamma_{j}} = \frac{\|\hat{\mathbf{U}}_{j}^{\mathrm{H}}\mathbf{H}_{j,j}\hat{\mathbf{V}}_{j}\|^{2}_{\mathrm{F}}}{\sum_{i \neq j} \|\hat{\mathbf{U}}_{j}^{\mathrm{H}}\mathbf{H}_{j,i}\hat{\mathbf{V}}_{i} \|^{2}_{\mathrm{F}} + \sigma_{\textrm{w}}^{{2}} \| \hat{\mathbf{V}}_{j} \|_{\mathrm{F}}^{2}}.
\end{equation} 
This \ac{SINR} is subsequently used to evaluate the achievable user throughput in the system. For the ideal case with perfect \ac{CSI}, the \ac{SINR} over all the streams of the $j$-th receiver, obtained using the transmit and receive matrices computed based on the actual channels, is expressed as~\cite{TF_IA_EUCNC}
\begin{equation} \label{eq: actual_SINR}
\gamma_{j} = \frac{\|\mathbf{U}_{j}^{\mathrm{H}}\mathbf{H}_{j,j}\mathbf{V}_{j}\|^{2}_{\mathrm{F}}}{\sum_{i \neq j} \|\mathbf{U}_{j}^{\mathrm{H}}\mathbf{H}_{j,i}\mathbf{V}_{j} \|^{2}_{\mathrm{F}} + \sigma_{\textrm{w}}^{{2}} \| \mathbf{V}_{j} \|_{\mathrm{F}}^{2}}.
\end{equation}
This value serves as a benchmark to compare the performance of the proposed \ac{TF-IA} approach.

\begin{figure}[t!]
    \centering
    \includegraphics[width=1\columnwidth,trim={0 0 0 0},clip]{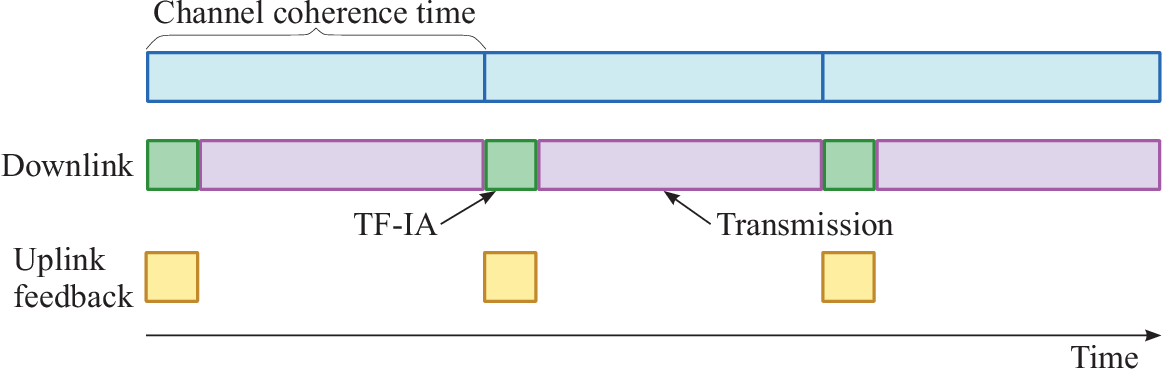}
    \caption{Timeline of the \ac{TF-IA} scheme.}
    \label{fig: TFIA_timeline}
\end{figure}

For simplicity, we implement the \ac{CSI} prediction as a sequence-to-one prediction task. Specifically, the proposed transformer model processes five input samples to predict the next time instant of the \ac{CSI} data. The modified transformer encoder consists of a multi-head attention block, layer normalization blocks, and a feed-forward block. Unlike the original transformer architecture~\cite{Attention_is_all_you_need}, we replace the fully connected layers in the position-wise feed-forward block with \ac{1D} \ac{CNN} layers. Additionally, the proposed model is equipped with a \ac{tanh} activation function instead of the \ac{ReLU}. The rationale for selecting \ac{1D} \ac{CNN} layers over fully connected layers is their superior ability to capture temporal correlations. In the downlink, within each channel coherence interval, the \ac{TF-IA} procedure and data transmission occur sequentially. The proposed \ac{TF-IA} scheme is summarized in Algorithm~\ref{algo: TF-IA}, with timeline illustrated in Fig.~\ref{fig: TFIA_timeline}.

To this end, we leverage the predicted \ac{CSI} obtained from the proposed transformer architecture within the \ac{IA} problem. However, directly incorporating these predictions into the \ac{IA} problem becomes impractical as the number of users grows large, due to the rapidly increasing computational complexity and the lack of tractable closed-form solution in large-dimensional networks. In the next section, we propose a model-free \ac{RL}-based approach that scales to large-scale \ac{MIMO} systems.

{\LinesNumberedHidden
\begin{algorithm}[!t]

    \textbf{Input:} Channel matrices $\mathbf{H}_{j, i}$, $\forall i,j=1, \dots ,K$;

    \textbf{Output:} Estimated $\hat{\mathbf{V}}_{j}$ and $\hat{\mathbf{U}}_{j}$, $\forall j=1, \dots ,K$;

\textbf{Define:} $L$, forecast horizon $H$, $D_{\textrm{m}}$, learning rate $\eta$, $h$, feedforward network dimension $f$ and number of encoders $N_{\textrm{enc}}$;

Reshape $\mathbf{H}_{j, i}$ using~\eqref{eq: reshaped_channel} and stack into input data $\mathbf{X}$;

\For{epoch $e \in \{1,\dots,E\}$}{
    \ForEach{training batch $(\mathbf{X}_{t-L+1:t}, \mathbf{Y}_{t+1:t+H}) $}{
        Perform input embedding using~\eqref{eq: input_embedding};
        
        Add positional encoding to embeddings;

        \For{$l \in \{1, \dots, N_{\textrm{enc}} \}$}{
              Compute multi-head self-attention using~(\ref{eq: attention})-(\ref{eq: head});
              
              Apply a residual connection followed by layer normalization, then a position-wise feed-forward network, and finally another residual normalization;
        }
      
        Predict $H$-step outputs $\hat{\mathbf{Y}}$;
        
        Compute the loss using~\eqref{eq: mse_loss};
        
        Update model parameters using Adam optimizer;
        
    }
}
Reshape $\hat{\mathbf{Y}}$ into channel matrix $\Tilde{\mathbf{H}}_{j, i}$;

Apply \ac{IA} with $\Tilde{\mathbf{H}}_{j, i}$ using~\eqref{eq: IA eq 1}--\eqref{eq: ith_column_of_Uj} and compute $\hat{\mathbf{V}}_{j}$ and $\hat{\mathbf{U}}_{j}$;

\textbf{Return} $\hat{\mathbf{V}}_{j}$, $\hat{\mathbf{U}}_{j}$, $\forall j=1, \dots ,K$;

\caption{Proposed \ac{TF-IA} algorithm}
\label{algo: TF-IA} \vspace{0mm}
\end{algorithm}}

 \begin{figure}[t!]
    \centering
    \includegraphics[width=0.9\columnwidth,trim={0 0 0 0},clip]{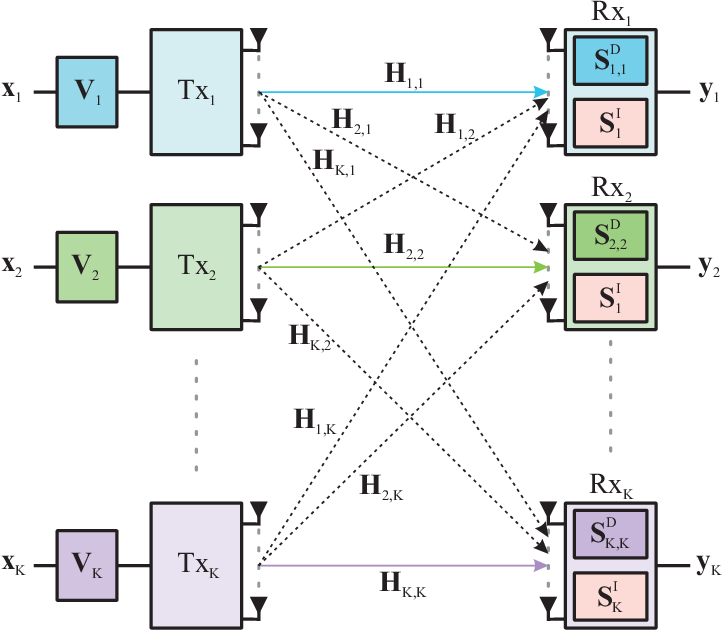}
    \caption{Subspace decomposition at the receiver under the proposed orthogonal alignment strategy.}
    \label{fig: subspace_decomposition_model}
\end{figure}

\section{RL-Based IA for Large-Scale MIMO} \label{sec: RL_IA}

In the previous section, we focused on mitigating the \ac{CSI} acquisition signaling overhead associated with \ac{IA} in small-scale \ac{MIMO} systems by predicting the \ac{CSI} using a transformer-based architecture. In this section, we complement the proposed model by leveraging a scalable solution that can overcome the computational burden of conventional \ac{IA} as the number of users grows large. To achieve scalability, we adopt a learning-based strategy rather than the traditional \ac{IA} algorithm. First, we reformulate the original sum throughput maximization problem $\mathbb{P}1$ as a distance minimization problem that surpasses the interference at the desired signal links. Then, we introduce the preliminaries of \ac{MDP} and \ac{DRL} needed to present the proposed \ac{RL-Blind} and \ac{RL-MaxSNR} schemes for large-scale \ac{MIMO} systems.

\subsection{Distance Minimization Problem} \label{subsec: problem_reformulation}

Consider the multi-user \ac{MIMO} system introduced in Section \ref{sec: multi_user_system_model}. At the receiver end, the received signal space is divided into two orthogonal subspaces~\cite{Inter_Cell_Interference_Sub_Space_Coordination} to separate the desired and interference signals, ensuring mutual orthogonality.
The key idea behind this separation is that desired signals are confined to the desired subspace, while interference signals are restricted to the interference subspace. The desired signal subspaces at the $j$-th receiver are denoted by $\mathbf{S}^{\textrm{D}}_{j}$ and $\mathbf{S}^{\textrm{I}}_{j}$, respectively. The precoding matrix of the $j$-th transmitter is designed to ensure that the desired signal $\mathbf{G}_{j,j}$ is projected onto $\mathbf{S}^{\textrm{D}}_{j}$, while directing the interference $\mathbf{G}_{i,j}$ into $\mathbf{S}^{\textrm{I}}_{i}$, $ \forall i \neq j$. Fig.~\ref{fig: subspace_decomposition_model} depicts the subspace decomposition at the receiver under the proposed orthogonal alignment strategy.
We introduce two distinct \ac{RL}-based strategies for subspace coordination, namely \ac{RL-Blind} and \ac{RL-MaxSNR}. The primary difference between these two strategies lies in how the desired and interference subspaces are declared and constructed. 
\begin{itemize}
    \item \textit{\ac{RL-Blind}: } The desired subspace $\mathbf{S}_{j}^{\textrm{D}}$ and the interference subspace $\mathbf{S}_{j}^{\textrm{I}}$ at the $j$-th receiver are defined arbitrarily. No strict policy governs these subspaces, except that they must remain orthonormal. Thus, these two subspaces serve as the desired and interference subspaces for each receiver in the multi-user \ac{MIMO} setting.

    \item \textit{\ac{RL-MaxSNR}: } In contrast to \ac{RL-Blind}, the \ac{RL-MaxSNR} method defines the desired and interference subspaces more systematically. For the $j$-th receiver, the desired subspace is obtained by performing eigendecomposition of $\mathbf{H}_{j,j}\mathbf{H}_{j,j}^{\mathrm{H}}$, ordering the eigenvalues in descending order, and selecting the eigenvectors corresponding to the $d_{j}$ largest eigenvalues. This eigenspace is then designated as the desired subspace of the $j$-th receiver, ensuring alignment of the desired signal with $\mathbf{S}_{j}^{\textrm{D}}$. The null space of this eigenspace is identified as the interference subspace.
\end{itemize}

Let us consider the $j$-th transmitter and define the chordal distance between the desired signal $\mathbf{G}_{j,j}$ and the desired subspace $\mathbf{S}_{j}^{\textrm{D}}$ as~\cite{Inter_Cell_Interference_Sub_Space_Coordination}
\begin{align}
        d(\mathbf{G}_{j,j}, \mathbf{S}^{\textrm{D}}_{j}) & = \frac{\|\mathbf{W}_{j,j} - \mathbf{W}^{\textrm{D}}_{j} \|_{\textrm{F}}}{\sqrt{2}}
        \\
        & = \bigg(\frac{1}{2} \textrm{Tr}\big[(\mathbf{W}_{j,j} - \mathbf{W}^{\textrm{D}}_{j})(\mathbf{W}_{j,j} - \mathbf{W}^{\textrm{D}}_{j})^{\mathrm{H}} \big]\bigg)^{\frac{1}{2}},
        \label{eq: chordal_distance}
\end{align}
where
\begin{equation}
        \mathbf{W}_{j,j} = \mathbf{G}_{j,j}(\mathbf{G}_{j,j}^{\mathrm{H}}\mathbf{G}_{j,j})^{-1}\mathbf{G}_{j,j}^{\mathrm{H}}
        \label{eq: orthogonal_projector_matrix}
\end{equation}
and $\mathbf{W}_{j}^{\textrm{D}}$ are the orthogonal projectors onto $\mathbf{G}_{j,j}$ and $\mathbf{S}^{\textrm{D}}_{j}$, respectively. The chordal distance $d(\mathbf{G}_{i,j}, \mathbf{S}_{i}^{\textrm{D}})$ between the interference signals from the $j$-th transmitter $\mathbf{G}_{i,j}$ and the interference subspace $\mathbf{S}_{i}^{\textrm{D}}$ is defined in a similar way. Minimizing the chordal distance is equivalent to maximizing the overlap between desired signal with interference signals. Therefore, using these distances, problem $\mathbb{P}1$ can be reformulated as a multi-objective distance minimization problem for the $j$-th transmitter as
\begin{mini!}|l|[2] 
{\mathbf{V}_{j}}{\delta_{j,j} d(\mathbf{G}_{j,j}, \mathbf{S}^{\textrm{D}}_{j}) - \sum_{j \neq i}\delta_{i, j} d(\mathbf{G}_{i,j}, \mathbf{S}^{\textrm{D}}_{i})}
{}{\mathbb{P}2: \quad}
{\label{eq: distance_minimization_objective}}
\addConstraint{\mathbf{V}^{\mathrm{H}}_{j} \mathbf{V}_{j} = \mathbf{I}_{d_{j}}, \quad \forall j \in \{1, \dots, K\}},
\end{mini!}
where $\delta_{i,j} = \frac{\alpha_{i,j}}{\sum_{i = 1}^{K}\alpha_{i,j}}$ represents the weight of each term in the objective function. These weights are derived from the path loss factors and are proportional to the relative strengths of the interference signals. By substituting \eqref{eq: chordal_distance} and \eqref{eq: orthogonal_projector_matrix} into \eqref{eq: distance_minimization_objective}, the objective function of problem $\mathbb{P}2$ can be further expanded as shown in \eqref{eq: expanded_P2} at the bottom of this page. As discussed in Section \ref{subsec: problem statement}, the complexity of the \ac{IA} scheme increases with the number of users, leading to a highly nonlinear optimization problem for which closed-form solutions are generally not available. Consequently, the reformulated problem $\mathbb{P}2$ is NP-hard and conventional optimization techniques are ineffective. To address this challenge, a scalable \ac{RL}-based \ac{IA} framework is proposed for large-scale \ac{MIMO} systems. Thus far, the throughput maximization problem has been reformulated as a distance minimization problem. The subsequent section introduces the \ac{MDP} formulation employed to solve problem $\mathbb{P}2$.

\begin{strip}
\hrulefill
\setcounter{equation}{19}
\begin{align}
    \mathbb{P}2 = \ & d_{j}(2\delta_{j,j} - 1) - \delta_{j,j} \: \text{Tr}\big[\mathbf{H}_{j,j}\mathbf{V}_{j} 
        (\mathbf{V}_{j}^{\mathrm{H}}\mathbf{H}_{j,j}^{\mathrm{H}}\mathbf{H}_{j,j}\mathbf{V}_{j})^{-1} 
        \mathbf{V}_{j}^{\mathrm{H}}\mathbf{H}_{j,j}^{\mathrm{H}} (\mathbf{W}_{j}^{\textrm{D}})^{\mathrm{H}}\big] \nonumber \\ & + \sum_{j \neq i}\delta_{i,j} \: \text{Tr}\big[\mathbf{H}_{i,j}\mathbf{V}_{j} 
        (\mathbf{V}_{j}^{\mathrm{H}}\mathbf{H}_{i,j}^{\mathrm{H}}\mathbf{H}_{i,j}\mathbf{V}_{j})^{-1} 
        \mathbf{V}_{j}^{\mathrm{H}}\mathbf{H}_{i,j}^{\mathrm{H}}(\mathbf{W}_{i}^{\textrm{D}})^{\mathrm{H}}\big].
    \label{eq: expanded_P2}
\end{align}
\setcounter{equation}{15}
\end{strip}

\subsection{Markov Decision Process Formulation} \label{subsec: MDP}
 
Given the dynamic, uncertain and sequential nature of wireless environments, the \ac{RL} problem is formulated via an \ac{MDP}, based on the Markov property. According to this property, the future state of a process depends only on the current state and is independent of past states. A finite \ac{MDP} is characterized by the tuple $\langle \mathcal{S}, \mathcal{A}, \mathcal{R}, \mathcal{P} \rangle$, where $\mathcal{S}$ is the set of states, $\mathcal{A}$ is the set of actions, $\mathcal{R}$ is the set of reward functions, and $\mathcal{P}$ is the transition probabilities at time instant $t$.
The agent observes the current state $s(t)$ at time $t$, takes an action $a(t)$, receives an immediate reward $r(t)$, and transits to the next state $s(t+1)$. 

Before presenting the solution approach, we first define the state space, action space, and reward function within the \ac{MDP} framework.
\begin{enumerate}
    \item \textit{State Space: } The state space represents all the necessary observations available to the agent at a given time instance. At time $t$, the $j$-th element of the state space $s_j(t)$ is defined as the distance of the $j$-th transmitter, obtained as
    \setcounter{equation}{20}
    \begin{equation} \label{eq: distance}
        s_j(t) = \delta_{j,j} \underbrace{ d(\mathbf{G}_{j,j}, \mathbf{S}^{\textrm{D}}_{j})}_{\text{Desired distance}} - \sum_{j \neq i}\delta_{i, j} \: \underbrace{d(\mathbf{G}_{i,j}, \mathbf{S}^{\textrm{D}}_{i}) }_{\substack{\text{Interference} \\ \text{distance}}}.
    \end{equation}
    Therefore, at time $t$, the state space can be expressed as
    \begin{equation} \label{eq: state_space}
        \mathcal{S}(t) =\{s_1(t), s_2(t), \dots , s_K(t)\}.
    \end{equation}

    \item \textit{Action Space: } The action space represents the set of possible decisions available to the agent at a given state. The $j$-th precoder can be written as $\mathbf{V}_{j} = \mathbf{A}_{j} + \text{i}\mathbf{B}_{j}$, with $\mathbf{A}_{j}, \mathbf{B}_{j} \in \mathbb{R}^{m_{j} \times d_{j}}$ and $\mathbf{V}^{\mathrm{H}}_{j} \mathbf{V}_{j} = \mathbf{I}_{d_{j}}, \forall j \in \{1, \dots, K\}$. At time $t$, the action space of the $j$-th transmitter is defined as 
    \begin{equation} \label{eq: action}
        \mathcal{A}_{j}(t) = \{ \mathbf{a}_{1}^{\mathrm{T}}, \dots, \mathbf{a}_{m_{j}}^{\mathrm{T}}, \mathbf{b}_{1}^{\mathrm{T}}, \dots, \mathbf{b}_{m_{j}}^{\mathrm{T}} \},
    \end{equation}
    where $\mathbf{a}_{k}$ and $\mathbf{b}_{k}$ denote the $k$-th columns of $\mathbf{A}_{j}$ and $\mathbf{B}_{j}$, respectively. Hence, at time $t$, the overall action space can be expressed as 
    \begin{equation} \label{eq: action_space}
        \mathcal{A}(t) =\{\mathcal{A}_{1}(t), \mathcal{A}_{2}(t), \dots , \mathcal{A}_{K}(t)\}.
    \end{equation}
    Following each actor output, the precoders are normalized to satisfy the constraint specified in $\mathbb{P}2$.

    \item \textit{Reward Function: } The reward function is defined as the weighted sum of the negative individual distances derived from the objective of problem $\mathbb{P}2$. The desired and interference subspaces are constructed to be mutually orthogonal. Hence, minimizing the distances of the desired and interference signals to their respective subspaces ensures that the \ac{IA} conditions are satisfied. Accordingly, at time $t$, the reward function is expressed as
    \begin{equation} \label{eq: reward_function}
        r(t) = \sum_{j = 1}^{K} \bigg( -\delta_{j,j} d(\mathbf{G}_{j,j}, \mathbf{S}^{\textrm{D}}_{j}) + \sum_{j \neq i}\delta_{i, j} d(\mathbf{G}_{i,j}, \mathbf{S}^{\textrm{D}}_{i}) \bigg).
    \end{equation}
\end{enumerate}

In an \ac{MDP}, the policy $\pi(s(t))$ defines the action selected in state $s(t)$, while the state transition probability and reward function are denoted by $p(s(t+1)\mid s(t), a(t))$ and $r(s(t), a(t))$, respectively. The optimal policy $\pi^{\star}$ maximizes the action-value function (Q-function), which can be defined as
\begin{equation} \label{eq: action_value1}
    Q_{\pi}(s(t), a(t)) = \mathbb{E}_{\pi} \left[ R(t) | s(t), a(t) \right],
\end{equation}
where $R(t) = \sum_{i=t}^{T} \gamma^{(i - t)} r(s(i), a(i))$ is the future reward with the discount factor $0 \leq \gamma \leq 1$, and $T$ denotes the episode horizon. The Q-function in \eqref{eq: action_value1} can be reformulated iteratively by incorporating the Bellman equation as
\begin{equation}
\label{eq: action_value2}
    Q_{\pi} \big( s(t), a(t) \big) = \mathbb{E}_{\pi} \! \left[ r(s(t), a(t)) + \gamma Q_{\pi} \big( s(t \! + \!1), a(t \! + \! 1) \big) \right].
\end{equation}

In real-world applications, high dimensionality poses significant challenges for traditional \ac{RL}, particularly in Q-learning. To address this issue, \ac{DRL} employs \ac{DNN} as function approximators for the Q-function and policies, enabling effective learning in high-dimensional state and action spaces. In \ac{MIMO} systems with \ac{IA}, the action space is continuous and both the state and action spaces have high dimension. To enable effective policy learning with improved exploration, a policy gradient approach is preferable. Therefore, we adopt \ac{DDPG}, as it efficiently handles continuous action spaces while leveraging deep function approximation for high-dimensional environments. The \ac{DDPG} approximates the deterministic policy using two \acp{DNN}, namely the actor and critic networks. In addition, target actor and critic networks are employed to improve the stability of the learning process. A replay buffer is used to store and randomly sample experiences collected during training~\cite{Continuous_control_with_DRL}. In the \ac{DDPG} framework, the actor and critic are parameterized by the learnable parameters $\theta^{\pi}$ and $\theta^{Q}$, respectively. In practice, both the actor and critic are modeled as neural networks. The critic $Q(s(t), a(t)\mid\theta^{Q})$ is learned using the Bellman equation, whereas the actor is updated using the policy gradient given by~\cite{Continuous_control_with_DRL}
\begin{align}
\label{eq: expected_return1}
    \nabla_{\theta^{\pi}} J
    &\approx \mathbb{E}\Big[
    \nabla_{\theta^{\pi}} Q(s,a|\theta^Q)
    \Big|_{s=s(t),a=\pi(s(t)|\theta^\pi)}
    \Big] \\
    &=
    \mathbb{E}\Big[
    \nabla_a Q(s,a|\theta^Q)
    \Big|_{s=s(t),a=\pi(s(t))}
    \Big. \nonumber\\
    &\hspace{3cm}
    \Big.
    \nabla_{\theta^\pi}\pi(s|\theta^\pi)
    \Big|_{s=s(t)}
    \Big].
\end{align}
The critic is updated using the loss
\begin{equation} \label{eq: mean_squared_belman_error}
    \mathcal{L}_{\textrm{DDPG}} = \mathbb{E} \left[ \left(Q\left(s(t), a(t) \right) - y(t) \right)^2 \right],
\end{equation}
with
\begin{equation} \label{eq: target}
    y(t) = r(t) + \gamma \max_{a(t+1)} Q\left(s(t+1), a(t+1) \right).
\end{equation}
The target networks are constructed as copies of the main networks (\emph{i.e.}, main actor and main critic). However, the target networks are periodically updated as
\begin{equation} \label{eq: soft_update1}
    \theta^{Q'} \leftarrow \tau \theta^{Q} + (1 - \tau) \theta^{Q'},
\end{equation}
\begin{equation} \label{eq: soft_update2}
    \theta^{\pi'} \leftarrow \tau \theta^{\pi} + (1 - \tau) \theta^{\pi'},
\end{equation}
where $\tau \ll 1$ denotes the soft-update factor that controls the target network update process, and $\theta^{\pi'}$ and $\theta^{Q'}$ represent the weights of the target actor and target critic networks, respectively. Algorithm~\ref{algo: RL-IA} summarizes the basic structure of the proposed \ac{DDPG}-based \ac{RL-MaxSNR} and \ac{RL-Blind} algorithms.

{\LinesNumberedHidden
\begin{algorithm}[!t]

    \textbf{Input:} Discount factor $\gamma$, learning rate $\eta$, $K$, $m_{\textrm{t}}$, $n_{\textrm{r}}$ $d_{j}$, $\mathbf{H}_{i,j}$;

    \textbf{Output:} Optimized $\mathbf{V}_{j}$, $\forall j=1, \dots ,K$;

Initialize replay buffer $\mathcal{D}$, actor network $\pi(s(t)|\theta^{\pi})$, critic network $Q(s(t), a(t) | \theta^{Q})$;

\For{episode $e \in \{1,\dots,E\}$}{

Randomly initialize precoding vectors $\mathbf{V}_j$;

\For{step $t \in \{1,\dots,T\}$}{

From the actor network, select action $a(t)$ according to current state;

Execute selected actions;

Observe next state $s(t+1)$ and compute reward $r(t)$;

Update the reply buffer $\mathcal{D}$;

Sample a mini-batch from $\mathcal{D}$; 

Compute the target value \eqref{eq: target};

Compute the loss using \eqref{eq: mean_squared_belman_error};

Update the target network using \eqref{eq: soft_update1} and \eqref{eq: soft_update2};

}
}
\textbf{Return} $\mathbf{V}_{j}$, $\forall j=1, \dots ,K$;
\caption{Structure of the proposed \ac{DDPG}-based \ac{RL-MaxSNR} and \ac{RL-Blind} algorithms.}
\label{algo: RL-IA} \vspace{0mm}
\end{algorithm}}
 \vspace{-1mm}

\section{Simulation results} \label{sec: simulation_results}

This section presents numerical results evaluating the effectiveness of the proposed \ac{TF-IA}, \ac{RL-Blind}, and \ac{RL-MaxSNR} schemes for \ac{IA}. We first report the \ac{CSI} prediction performance, followed by the results of the proposed \ac{RL}-based subspace coordination. Prior to performance evaluation, we describe the simulation data generation process and define the baseline schemes for each approach. We use two hidden layers in both the actor and critic networks of the \ac{RL} models, with $h_{1}$ and $h_{2}$ neurons, respectively. All simulations of the proposed algorithms are performed on a single NVIDIA Tesla V100 GPU using the PyTorch framework. The simulation parameters for the transformer model of \ac{TF-IA}, \ac{RL-Blind}, and \ac{RL-MaxSNR} are summarized in Table~\ref{table: hyper-params}.

\subsection{Simulated Data Generation} \label{sec: simulated_data_generation}

In real-world scenarios, channels exhibit temporal correlations with previous \ac{CSI} data, which is crucial for accurate \ac{CSI} prediction. To simulate this, we first generate time uncorrelated Rayleigh faded channels, each containing $T$ time samples, and then apply a \ac{FIR} filter to introduce temporal correlations into the generated channels. The generated correlated channel data are feature-scaled using min-max normalization to improve convergence during transformer training and validation. The data set contains a total of $20 \times 10^3$ samples, divided into $60\%$ for training, $20\%$ for validation, and $20\%$ for testing the transformer model. Although time-varying \ac{CSI} scenarios exist, this study considers that the variation is slow enough that the trained transformer model remains valid for a significant duration. Such an assumption is valid, for example, in indoor scenarios with limited mobility.

\begin{table}[t!]
\centering
 \caption{Simulation parameters}
\label{table: hyper-params}
\begin{tabular}{@{}ll|ll@{}}
\toprule
\multicolumn{2}{c|}{\textbf{Transformer model}} & \multicolumn{2}{c}{\textbf{\ac{RL-Blind} \& \ac{RL-MaxSNR} }} \\
\midrule
\textbf{Parameter} & \textbf{Value} & \textbf{Parameter} & \textbf{Value} \\ 
\midrule
$L$ & $5$ & Hidden layers & $2$ \\
$f_{\textrm{in}}$ & $72$ & $h_{1}$ & $400$ \\
$D_{\textrm{m}}$ & $120$ & $h_{2}$ & $300$ \\
$h$ & $2$ & Replay memory & $1 \times 10^{6}$ \\
$N_{\textrm{enc}}$ & $2$ & $\gamma$ & $0.99$ \\
$f$ & $144$ &  Critic learning rate & $1 \times 10^{-3}$ \\
Activation & $\textrm{tanh}$ &  Activation & ReLU \\
Optimizer & Adam & Optimizer & Adam \\
Loss function & MSE &  State transitions & $20$ \\
Epochs & $100$ & Episodes & $100$ \\
\bottomrule
\end{tabular}
\end{table}

\subsection{Baseline Schemes} \label{subsec: baseline_schemes}

We compare the proposed \ac{TF-IA} algorithm with established time series models, namely \ac{LSTM-IA} and \ac{CNN-IA}, which have been validated in~\cite{CSI_pred-2} and~\cite{CSI_pred-1}, respectively. The proposed \ac{RL-Blind} and \ac{RL-MaxSNR} algorithms are compared against four baseline schemes:
\begin{itemize}
    \item Blind: The precoding matrices are constructed from randomly generated orthonormal vectors;
    \item  MaxSNR: Each transmitter independently selects its precoding matrix to maximize the received \ac{SNR} at its corresponding receiver;
    \item \Ac{Blind-ICISC}: The precoding matrices are conventionally designed using randomly generated desired subspaces~\cite{Inter_Cell_Interference_Sub_Space_Coordination};
    \item  \Ac{MaxSNR-ICISC}: The precoding matrices are conventionally designed based on the eigenmodes of the channel matrices~\cite{Inter_Cell_Interference_Sub_Space_Coordination}.
\end{itemize}
Furthermore, the three proposed algorithms are evaluated against the distributed \ac{IA} method in~\cite{Distributed_Numerical_Approach_to_IA} as a reduced-feedback baseline.

\subsection{TF-IA Results and Analysis} \label{subsec: TF_IA_results}

In the simulations, we consider small-scale \ac{MIMO} networks, for which the proposed \ac{TF-IA} approach has been specifically designed, as discussed in Section~\ref{sec: TF_IA}. Following the system model in Section~\ref{subsec: multi user system model}, we assume $K = 3$, $m_{\textrm{t}} = 2$, $n_{\textrm{r}} = 2$, and $d_{j} = 1$, $\forall j = 1, \dots, K$. In this study, the \ac{CSI} prediction problem is modeled as a multivariate time series forecasting problem. For the performance evaluation between proposed transformer-based prediction and conventional \ac{LSTM} and \ac{CNN}-based methods, we use \ac{NMSE} as the performance metric, which is expressed as
\begin{equation}
\label{eq: NMSE}
    \textrm{NMSE}(\mathbf{H}_{j,i}, \Tilde{\mathbf{H}}_{j,i}) = 10\log_{10}\bigg( \mathbb{E}\bigg[ \frac{\| \mathbf{H}_{j,i} - \Tilde{\mathbf{H}}_{j,i} \|^{2}_{\textrm{F}}}{\| \mathbf{H}_{j,i} \|^{2}_{\textrm{F}}} \bigg] \bigg),
\end{equation}
where $\Tilde{\mathbf{H}}_{j,i}$ denotes the predicted channel matrix. Table~\ref{table: NMSE} presents the \ac{NMSE} corresponding to each prediction model used in this study. The \ac{NMSE} of the prediction of the proposed transformer model is orders of magnitude lower than both \ac{LSTM} and \ac{CNN}, which reflects better convergence. This gain is mainly due to the transformer's attention mechanism, which effectively handles complex scenarios involving long sequences, especially given the high dimensionality of input features in this study. Furthermore, transformer models are highly effective at capturing global information, making them well-suited for the CSI prediction task. We further observe that the \ac{NMSE} is equally low in the overall considered \ac{SNR} regime. 

\begin{table}[t!]
\centering
 \caption{Summary of the \ac{NMSE} for each \ac{CSI} prediction architecture for different \ac{SNR} values}
\label{table: NMSE}
\begin{tabular}{@{}lcccccc@{}}
\toprule
\multirow{2}{*}{\textbf{Architecture}} & \multicolumn{6}{c}{\textbf{NMSE (dB) for \ac{SNR} values}} \\
\cmidrule{2-7}
& $0 \: \textrm{dB}$ & $5 \: \textrm{dB}$ & $10 \: \textrm{dB}$ & $15 \: \textrm{dB}$ & $20 \: \textrm{dB}$ & $25 \: \textrm{dB}$\\ 
\midrule
\ac{LSTM} & $-1.25$ & $-0.66$ & $-0.25$ & $-0.69$ & $-1.69$ & $-1.46$ \\
\hline
\ac{CNN} & $0.54$ & $0.46$ & $0.96$ & $0.28$ & $-2.23$ & $-0.68$  \\
\hline
\textbf{Transformers} & $-14.32$ & $15.92$ & $-16.12$ & $-15.55$ & $-16.24$ & $-16.56$  \\
\bottomrule
\end{tabular}
\end{table}

\begin{figure}[t!]
    \centering
    \includegraphics[width=1\columnwidth,trim={0 0 0 10pt},clip]{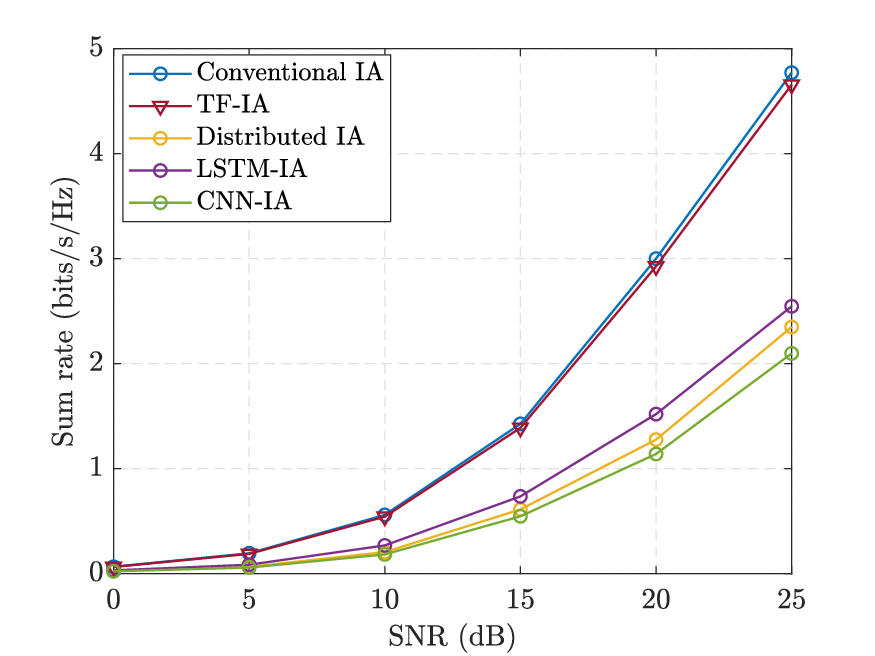}
    \caption{Variation of sum rate with \ac{SNR} for a \ac{MIMO} system with $K = 3$, $m_{\textrm{t}} = n_{\textrm{r}} = 2$, and $d_j = 1$, with the desired links transmitted at higher power than the interfering links.}
    \label{fig: sum_rate_diff}
\end{figure}

In this experiment, we evaluate the impact of \ac{CSI} prediction on the performance of the \ac{IA} algorithm. Following Section~\ref{sec: multi_user_system_model}, the small-scale \ac{MIMO} system is configured with the aforementioned parameters. The proposed \ac{TF-IA} approach is examined under two scenarios. In the first scenario, the transmit power of the desired links varies from $0$ dB to $25$ dB, while the interfering link power is modeled as $P \sim \mathcal{U}(\text{SNR}$-10$, \text{SNR}$-15$)$. This represents a weak interference scenario where the interference signals are $10$--$15$~dB weaker than the mean desired signal strength. Fig.~\ref{fig: sum_rate_diff} presents the variation in sum rate performance across the given \ac{SNR} range, with the desired links transmitted at higher power than the interfering links. As shown, the proposed \ac{TF-IA} closely follows the sum rate performance of the conventional \ac{IA} algorithm, whereas the performance of \ac{LSTM-IA}, \ac{CNN-IA}, and distributed \ac{IA} is comparatively lower. Across the evaluated \ac{SNR} range, \ac{TF-IA} consistently outperforms all three baseline algorithms in both low- and high-\ac{SNR} regions. The proposed \ac{TF-IA} achieves up to $97.27\%$ of the sum rate performance of the conventional \ac{IA} algorithm, whereas the baseline methods \ac{LSTM-IA}, \ac{CNN-IA}, and distributed \ac{IA} achieve only $49.32\%$, $36.26\%$, and $40.62\%$, respectively.

\begin{figure}[t!]
    \centering
    \includegraphics[width=1\columnwidth,trim={0 0 0 10pt},clip]{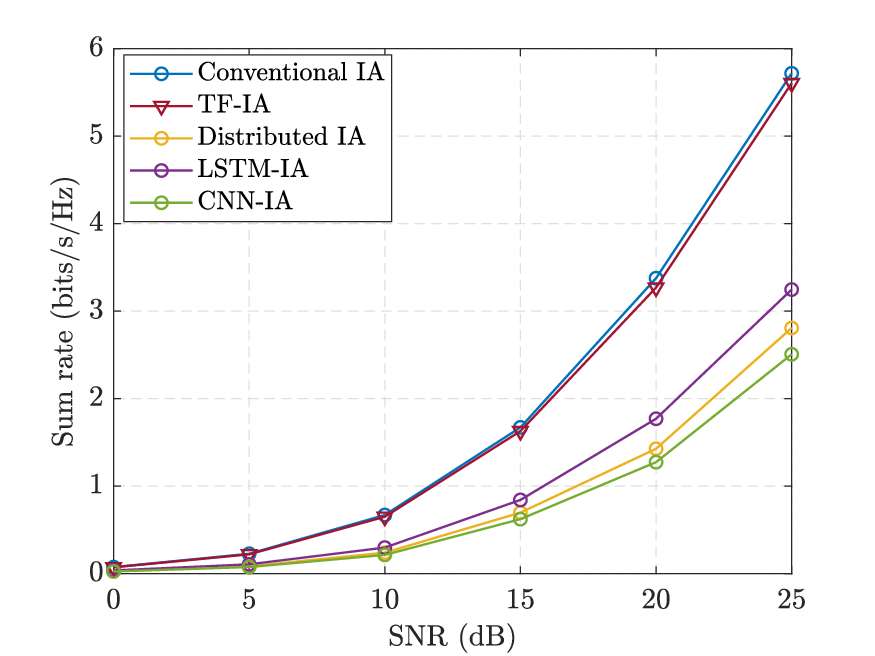}
    \caption{Variation of sum rate with \ac{SNR} for a \ac{MIMO} system with $K = 3$, $m_{\textrm{t}} = n_{\textrm{r}} = 2$, and $d_j = 1$, with equal transmit power for both desired and interfering links.}
    \label{fig: sum_rate_same}
\end{figure}

To gain further insight, as the second scenario, we examine the sum rate variation when the desired and interfering links transmit with equal power in the same \ac{MIMO} system across the given \ac{SNR} range, reflecting a strong interference scenario. Fig.~\ref{fig: sum_rate_same} illustrates the variation of the sum rate under equal transmit power for both desired and interfering links. In this scenario, since the transmit powers of the desired and interfering signals are identical, recovering the desired signal becomes more challenging than in the first scenario. However, the proposed \ac{TF-IA} consistently achieves the highest sum rate among the baseline methods and closely follows the performance of the conventional \ac{IA} algorithm. Compared to the first scenario, the sum rate performance decreases when the desired and interfering links transmit with identical power. Nevertheless, the proposed \ac{TF-IA} approach demonstrates superior performance, achieving near-optimal results that closely align with the conventional \ac{IA} algorithm.

When the transmit power is identical for all links, the proposed \ac{TF-IA} achieves up to $96.24\%$ of the sum rate attained by the conventional \ac{IA} algorithm, whereas \ac{LSTM-IA}, \ac{CNN-IA}, and distributed \ac{IA} achieve only $50.16\%$, $36.41\%$, and $40.78\%$, respectively. Overall, the sum rate variation and \ac{NMSE} comparison clearly demonstrate that the proposed \ac{TF-IA} method outperforms all three baseline approaches.

\subsection{RL-Blind and RL-MaxSNR Results and Analysis} \label{subsec: RL results}

This section discusses the training performance of the \ac{DDPG} model and provides a comprehensive analysis on \ac{RL-Blind} and \ac{RL-MaxSNR} across different \ac{SNR} levels and \ac{MIMO} architectures, especially large-scale architectures.

\begin{figure}[t!]
    \centering
    \includegraphics[width=1\columnwidth,trim={0 0 0 10pt},clip]{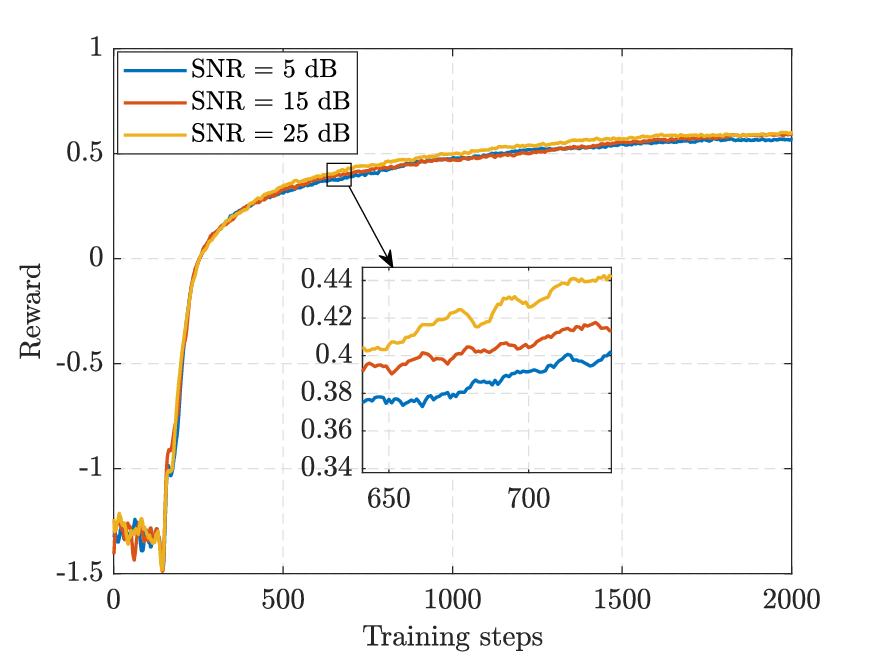}
    \caption{Convergence of the proposed \ac{RL-MaxSNR} approach for a single transmitter using the \ac{DDPG} model as a function of training steps.}
    \label{fig: reward_variation}
\end{figure}

\begin{figure}[t!]
    \centering
    \begin{subfigure}{1\linewidth}
        \centering
        \includegraphics[width=1\columnwidth,trim={0 18pt 0 35pt},clip]{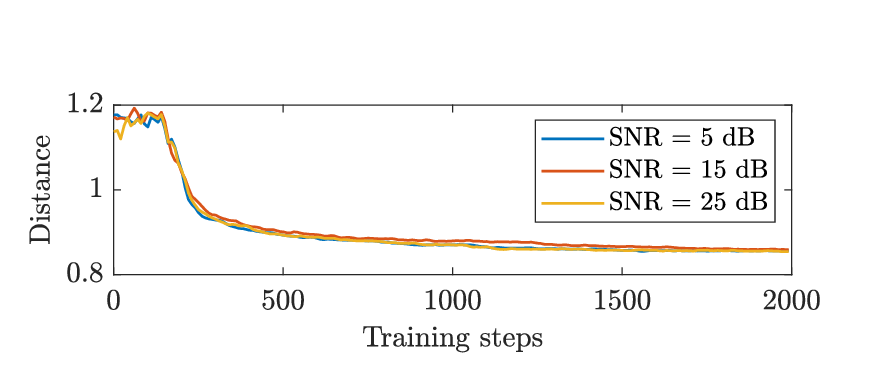}
        \caption{Desired signal distance.}
        \label{subfig: desired_distance}
    \end{subfigure}
    
    \begin{subfigure}{1\linewidth}
        \centering
        \includegraphics[width=1\columnwidth,trim={0 18 0 35pt},clip]{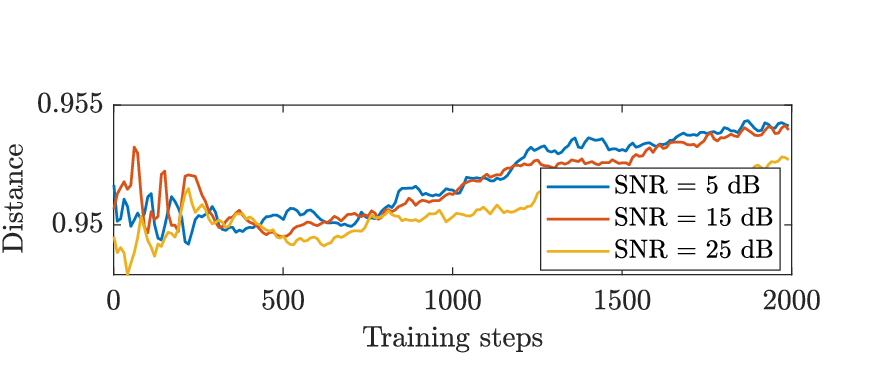}
        \caption{Interference signal distance.}
        \label{subfig: interference_distance}
    \end{subfigure}

    \begin{subfigure}{1\linewidth}
        \centering
        \includegraphics[width=1\columnwidth,trim={0 18pt 0 35pt},clip]{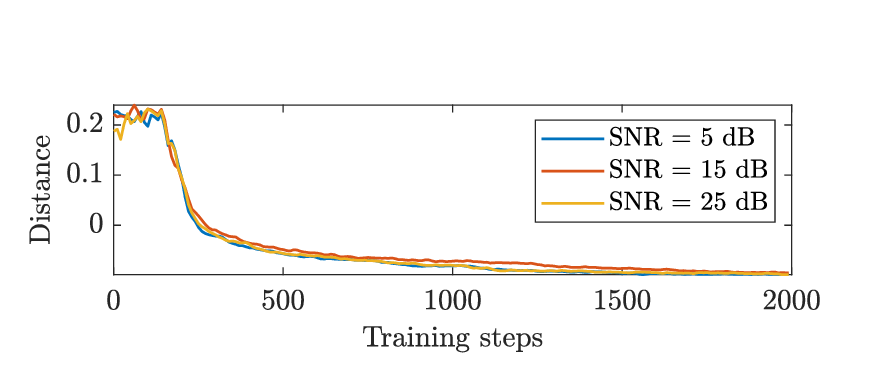}
        \caption{Total distance.}
        \label{subfig: total_distance}
    \end{subfigure}

    \caption{Distance variation of the proposed RL-MaxSNR model as a function of training steps: (a) Desired signal distance variation, (b) interference signal distance variation, and (c) total distance variation.}
    \label{fig: distance_variation}
\end{figure}

\subsubsection{Performance of the proposed \ac{RL-MaxSNR} approach}
Fig.~\ref{fig: reward_variation} shows the convergence behavior of the proposed \ac{RL-MaxSNR} approach using the \ac{DDPG} model as a function of training steps for \ac{SNR} values of 5, 15, and 25 dB. The multi-user \ac{MIMO} system is configured with six users, each having four transmit and receive antennas transmitting a single data stream, \emph{i.e.}, $K = 6$, $m_{\textrm{t}} = 4$, $n_{\textrm{r}} = 4$, and $d_{j} = 1$. The reward function in \eqref{eq: reward_function} is formulated to promote stable convergence. At the beginning of training, all the rewards are at their minimum. As the training progresses, the rewards increase and eventually saturate within their valid margin. 

Fig.~\ref{fig: distance_variation} depicts the individual and total distance variation of the proposed \ac{RL-MaxSNR} model as a function of the training steps. The weighted desired distance $\delta_{j,j}d(\mathbf{S}_{j,j}^{\textrm{C}}, \mathbf{S}^{\textrm{D}}_{j})$ is minimized, while the weighted sum of the interference distance $\sum_{j \neq i}\delta_{i, j} d(\mathbf{S}_{i,j}^{\textrm{C}}, \mathbf{S}^{\textrm{D}}_{i})$ is maximized. The reduction in total distance observed in Fig.~\ref{fig: distance_variation} verifies the consistency of the proposed model with \eqref{eq: distance}.

\begin{figure}[t!]
    \centering
    \includegraphics[width=1\columnwidth,trim={0 0 0 10pt},clip]{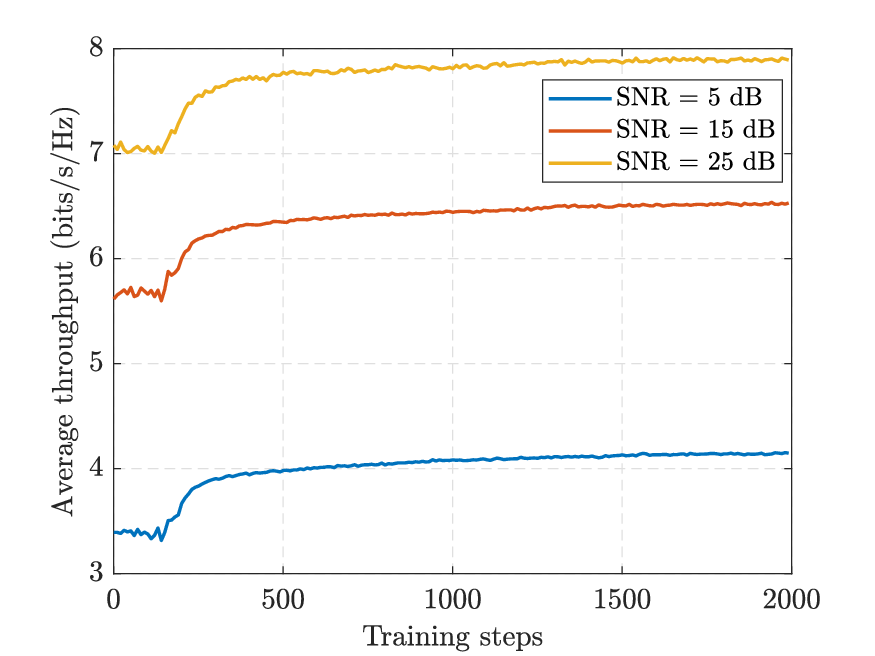}
    \caption{Convergence of the throughput of proposed \ac{RL-MaxSNR} as a function of training steps.}
    \label{fig: throughput_variation}
\end{figure}

Fig.~\ref{fig: throughput_variation} demonstrates the convergence of the average throughput of the proposed \ac{RL-MaxSNR} as a function of training steps. When the desired signals are aligned with the desired subspaces, the throughput of each receiver is maximized. As the reward functions are optimized, the average user throughput also increases and reaches its maximum, as shown in Fig.~\ref{fig: throughput_variation}. It may be noted that the same reward (\emph{i.e.}, alignment of the desired and interference subspaces) leads to different average throughput under different average \ac{SNR} values. The \ac{RL-Blind} scheme exhibits a similar performance trend, and is therefore omitted for the ease of presentation.

\subsubsection{Performance evaluation across different experimental scenarios}
\begin{figure}[t!]
    \centering
    \includegraphics[width=1.1\columnwidth,trim={0 0 0 10pt},clip]{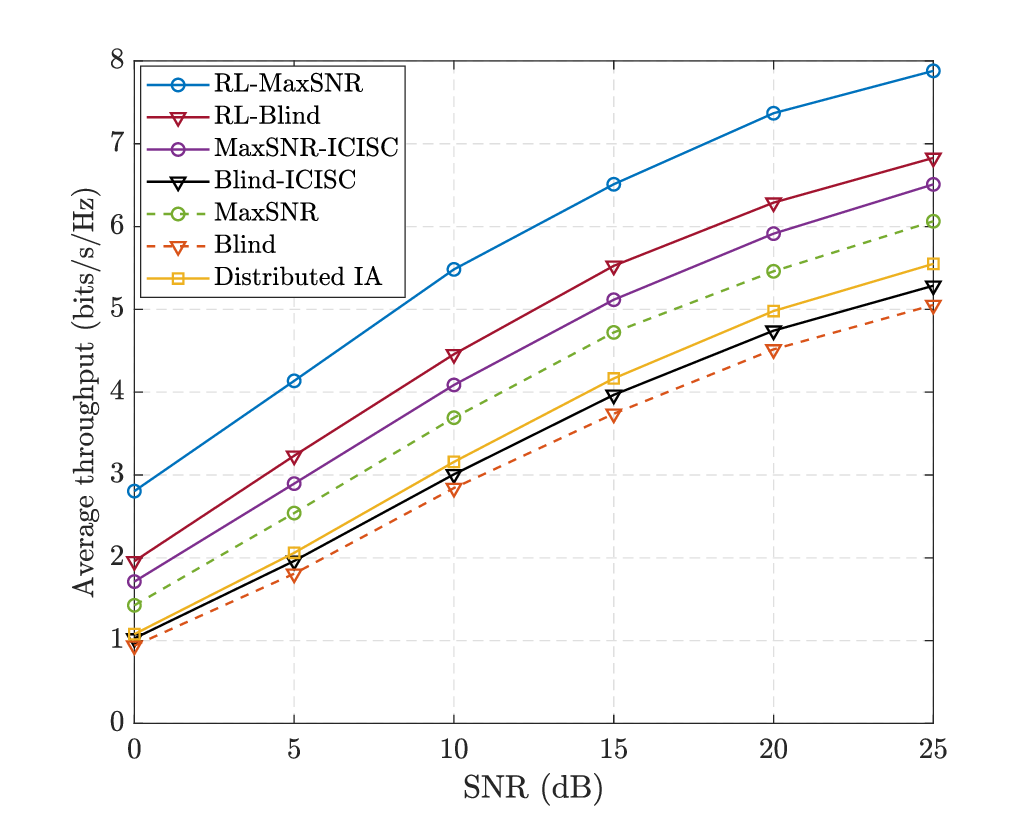}
    \caption{Variation of average user throughput with \ac{SNR} for a \ac{MIMO} system with $K = 6$, $m_{\textrm{t}} = n_{\textrm{r}} = 4$, and $d_j = 1$, with the desired links transmitted at higher power than the interfering links.}
    \label{fig: different_snr}
\end{figure}

Next, we evaluate the performance of the proposed \ac{RL}-based subspace coordination methods across three experimental scenarios. In the first scenario, the multi-user \ac{MIMO} system is configured with $K=6$, $m_{\textrm{t}} = 4$, $n_{\textrm{r}} = 4$, and $d_{j} = 1$. The desired link transmit power ranges from $0$ dB to $25$ dB, while the interfering link power is modeled as $P \sim \mathcal{U}(\text{SNR}$-10$, \text{SNR}$-15$)$ corresponding to weak interference conditions. The performance of \ac{RL-Blind} and \ac{RL-MaxSNR} is compared against the baseline schemes described in Section~\ref{subsec: baseline_schemes}. Fig.~\ref{fig: different_snr} depicts the average user throughput as a function of the \ac{SNR}. The proposed approaches consistently surpass all baseline schemes across the entire \ac{SNR} range. In particular, \ac{RL-MaxSNR} provides a clear performance gain over \ac{RL-Blind}, highlighting the importance of defining meaningful subspaces to align both desired and interference signals at the receiver. Although the desired signal power exceeds the interference power in this scenario, the conventional baseline schemes still fail to achieve the average throughput attained by \ac{RL-MaxSNR}. Numerically, at $25$ dB, \ac{RL-MaxSNR} yields a $17.5 \%$ average throughput gain, and \ac{RL-Blind} achieves a $4.8 \%$ improvement over \ac{MaxSNR-ICISC}, the strongest baseline. These results demonstrate the clear advantage of RL-based approaches over conventional schemes.

\begin{figure}[t!]
    \centering
    \includegraphics[width=1.1\columnwidth,trim={0 0 0 10pt},clip]{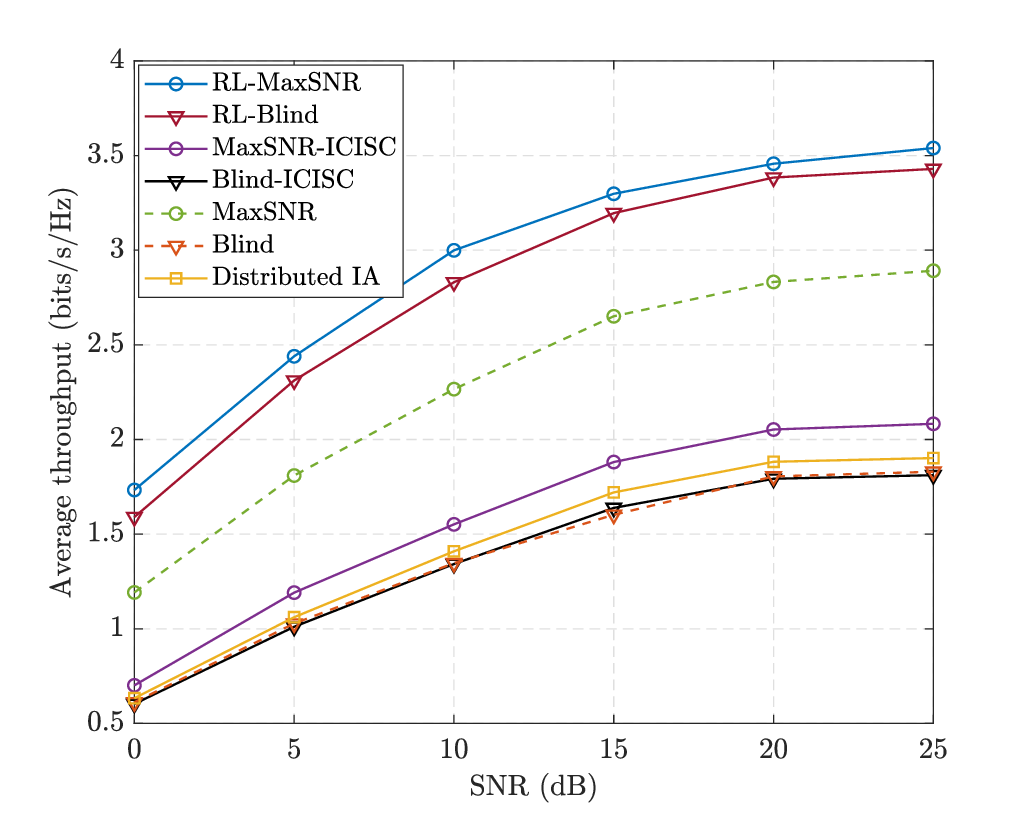}
    \caption{Variation of average user throughput with \ac{SNR} for a \ac{MIMO} system with $K = 6$, $m_{\textrm{t}} = n_{\textrm{r}} = 4$, and $d_j = 1$, with equal transmit power for both desired and interfering links.}
    \label{fig: same_snr}
\end{figure}

In the second scenario, the same \ac{MIMO} configuration and \ac{SNR} range are maintained. However, the transmit power of the desired and interference links is modeled with identical \ac{SNR} values reflecting a strong interference scenario. This setting creates a more challenging interference environment compared to the first scenario, as the desired signals no longer dominate in power. Fig.~\ref{fig: same_snr} shows the average user throughput as a function of the \ac{SNR}, where desired and interference links have equal transmit power. Both \ac{RL-MaxSNR} and \ac{RL-Blind} still outperform all baseline schemes, highlighting the ability of RL-based approaches to maintain high performance even in a more challenging interference-limited environment. The achieved throughput as function of \ac{SNR} is lower compared to the first scenario, as the desired and interference signal powers are equal. At $25$ dB, \ac{RL-MaxSNR} and \ac{RL-Blind} achieve average user throughput gains of $29.13\%$ and $25.92\%$, respectively, over the best-performing baseline (MaxSNR). These results underscore the effectiveness of RL-based approaches in maintaining high performance under interference-limited conditions.

\begin{figure}[t!]
    \centering
    \includegraphics[width=1.1\columnwidth,trim={0 0 0 10pt},clip]{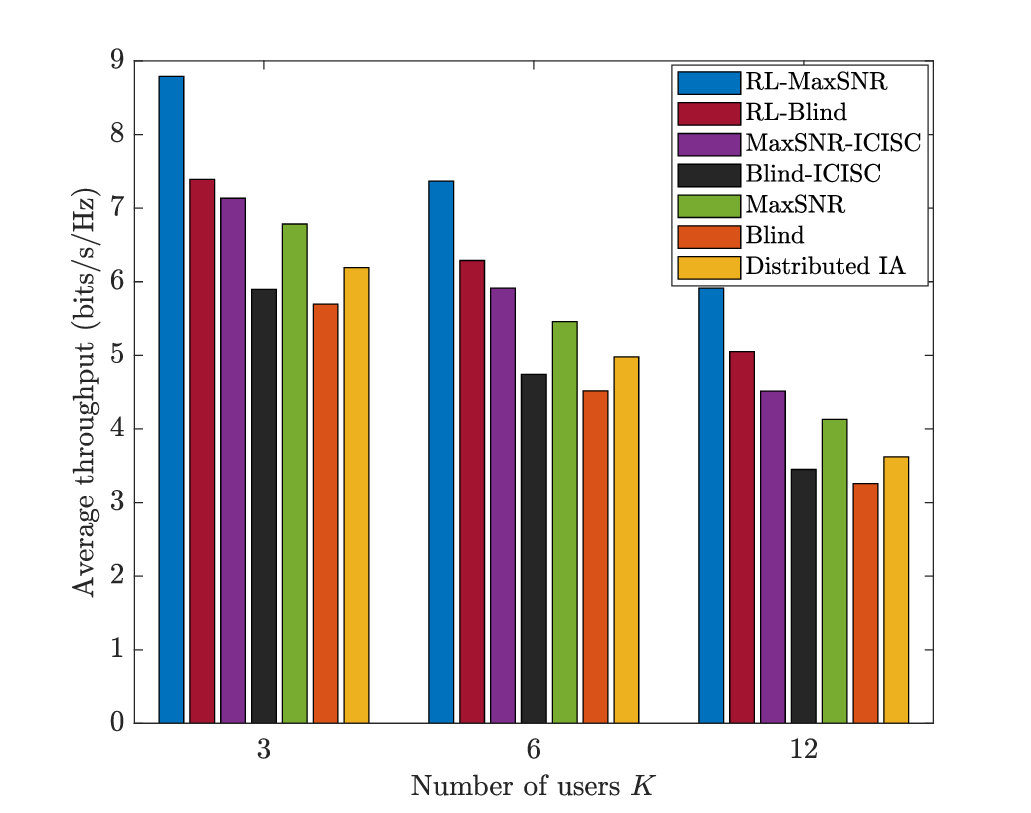}
    \caption{Impact of the proposed schemes on average user throughput across varying user counts in different \ac{MIMO} systems at $\text{SNR} = 20~\text{dB}$, with equal transmit power for both desired and interfering links.}
    \label{fig: multiple_users}
\end{figure}

In the third scenario, we examine the effect of the number of users on the throughput. Specifically, we use a fixed \ac{SNR} of $20$~dB, while the $K$ users in the \ac{MIMO} configuration is varied as $3$, $6$, and $12$. Fig.~\ref{fig: multiple_users} presents the average user throughput of each scheme as a function of the number of users. As the user count increases, the average user throughput decreases. Although all schemes exhibit this trend, the proposed \ac{RL-MaxSNR} and \ac{RL-Blind} achieve consistently higher average user throughput and significantly outperform all baseline methods across the evaluated user counts.

\subsection{Computational Complexity Analysis}
We investigate the computational complexity of the proposed \ac{TF-IA}, \ac{RL-Blind}, and \ac{RL-MaxSNR} schemes in terms of \ac{FLOPs}. The complexity of the transformer-based \ac{IA} framework is given by $\mathcal{C}(\textrm{TF-IA}) = 2Lf_{\textrm{in}}D_{\textrm{m}} + N_{\textrm{enc}}(8LD_{\textrm{m}}^{2} + 4L^{2}D_{\textrm{m}} + 2L(fD_{\textrm{m}}k_{1} + D_{\textrm{m}}fk_{2}) + 2D_{\textrm{m}}f_{\textrm{in}}$, where $k_{1}$ and $k_{2}$ denote the kernel sizes of the two \ac{CNN} layers~\cite{complexity_analysis_TF_IoT_J}.
For both \ac{RL-Blind} and \ac{RL-MaxSNR}, the same \ac{DDPG} model is used, and the computational complexity is determined by the actor-critic networks. The total \ac{FLOPs} per forward pass can be expressed as $\mathcal{C}(\textrm{RL-Blind}), \mathcal{C}(\textrm{RL-MaxSNR})  = 2 \big( h_{1}(2 |\mathcal{S}| + |\mathcal{A}|) + 2h_{1}h_{2} + h_{2}(|\mathcal{A}| + 1)\big)$, where $|\mathcal{S}|$ and $|\mathcal{A}|$ denote the state and action dimensions, respectively. Finally, sample efficiency is quantified by the number of training samples required for \ac{TF-IA} and the number of environment transitions required for the two \ac{RL}-based approaches to reach $95\%$ of the average sum rate. The training time ($T_{\textrm{train}}$), inference time ($T_{\textrm{infer}}$), \ac{FLOPs} count, and sample efficiency of each model are summarized in Table~\ref{table: Complexity}. The training time of \ac{TF-IA} is significantly higher than that of the two \ac{RL}-based approaches. Similarly, the inference time of \ac{TF-IA} is slightly higher than that of the \ac{RL}-based approaches. The total number of \ac{FLOPs} follows a similar trend, indicating the higher computational cost of the \ac{TF-IA} model. However, in terms of sample efficiency, \ac{TF-IA} and the two \ac{RL}-based approaches exhibit different behaviors, and a direct comparison may not be entirely fair due to the differences in their model architectures.

\begin{table}[t!] 
\centering
\caption{Summary of the training time, inference time, \ac{FLOPs} count, and sample efficiency of the proposed algorithms}
\label{table: Complexity}
\begin{tabular}{@{}lcccl@{}}
\toprule
\textbf{Algorithm} & $\textbf{\textit{T}}_{\textrm{train}}$ & $\textbf{\textit{T}}_{\textrm{infer}}$ & \textbf{\ac{FLOPs}} & \textbf{Sample efficiency}\\ 
\midrule
\multirow{2}{*}{\ac{TF-IA}} & \multirow{2}{*}{$3.65$ min} & \multirow{2}{*}{$0.90$ ms} & \multirow{2}{*}{$1.99 \times 10^{6}$} & $5.76 \times 10^{5}$ training \\
 & & & & samples \\
\midrule
\ac{RL-Blind} \& & \multirow{2}{*}{$45$ s} & \multirow{2}{*}{$0.30$ ms} & \multirow{2}{*}{$5.57 \times 10^{5}$}  & $1500$ environment \\
 \ac{RL-MaxSNR} & & & & transitions \\
\bottomrule
\end{tabular}
\end{table}

\subsection{Advantages and Potential Limitations of the Proposed Algorithms}

Achieving \ac{DoF} optimal interference management with \ac{IA} typically requires global \ac{CSI} at the transmitter. Since \ac{TF-IA} does not rely on channel estimation, it eliminates the feedback overhead, achieving $100\%$ feedback bit savings compared to conventional \ac{IA}. In \ac{RL-Blind}, the subspaces are determined independently of channel information, eliminating the need for receivers to communicate subspace information to the transmitters. Nevertheless, channel knowledge is still required during both training and deployment to learn and execute the subspace coordination policy. As a result, \ac{RL-Blind} achieves a $50\%$ reduction in feedback overhead compared with traditional \ac{IA}. In contrast, the desired subspace computation in \ac{RL-MaxSNR} is channel-dependent and therefore requires receiver feedback to update the subspace as the channel varies. In terms of pilot usage, \ac{TF-IA} eliminates pilot signaling entirely, whereas \ac{RL-Blind} and \ac{RL-MaxSNR} do not reduce pilot usage relative to conventional \ac{IA}.

The \ac{TF-IA} approach reduces pilot usage by $100\%$ but incurs higher computational complexity, as it relies on learning from historical channel data. In contrast, \ac{RL-MaxSNR} has lower computational complexity but requires explicit channel information to derive the desired subspaces, resulting in increased signaling overhead. Furthermore, \ac{RL-Blind} achieves improved performance with lower computational complexity compared to \ac{TF-IA} in large-scale \ac{MIMO} systems. Therefore, the selection among these approaches depends on system design priorities, particularly the trade-off between computational complexity, signaling overhead, and the scale of the \ac{MIMO} system.

\section{Conclusion} \label{sec: Conclusion}
In this study, we investigated the practical challenges of applying \ac{IA} in downlink multi-user \ac{MIMO} systems. We identified and analyzed key parameters that can be optimized to enable efficient \ac{IA}. We introduced \ac{TF-IA}, which employs \ac{CSI} prediction to reduce signaling overhead in small-scale \ac{MIMO} systems, and demonstrated its superior performance compared to three considered baseline methods. Moreover, we formulated a multi-objective optimization problem and proposed two \ac{RL}-based approaches, \ac{RL-Blind} and \ac{RL-MaxSNR}, to address the scalability issue in large-scale \ac{MIMO} systems, achieving effective interference management and improved system performance. The performance of the proposed \ac{RL-MaxSNR} and \ac{RL-Blind} schemes were investigated under varying numbers of users and interference conditions, benchmarking them against conventional approaches. Simulation results indicated that at $25$ dB, \ac{RL-MaxSNR} and \ac{RL-Blind} achieve average user throughput gains of $29.13\%$ and $25.92\%$, respectively, over the best-performing baseline, highlighting the effectiveness of the proposed approaches in managing interference and enhancing system performance. Extending the proposed \ac{RL}-based framework to \ac{MARL} for \ac{IA} optimization is left for future work.

\bibliographystyle{IEEEtran}
\bibliography{General_resources/references}

\end{document}